\documentclass{aa} 
\usepackage{natbib}
\usepackage{graphicx}
\usepackage{txfonts}
\usepackage{booktabs}
\usepackage{multirow}
\usepackage{lipsum}
\usepackage{subcaption}         
\usepackage{lscape}             
\usepackage{placeins}           
                                
\usepackage{svg}
\usepackage{tablefootnote}
\usepackage{import}
\usepackage{tikz}
\input{texplot/layers/Ball.sty}
\input{texplot/layers/Box.sty}
\input{texplot/layers/RightBandedBox.sty}
\usetikzlibrary{positioning,3d,quotes,arrows.meta}
            
\usepackage[dvipsnames]{xcolor}
\usepackage[colorlinks=true, allcolors=blue]{hyperref}
\usepackage{comment}
\titlerunning{Interpretable Human-Label-Free Real-Bogus with UQ}
\authorrunning{Bonnet-Guerrini et al.}
\begin{document}

   \title{Interpretable Human-Label-Free Deep Learning for Real-Bogus Classification with Uncertainty Quantification}



   \author{Rapha{\"e}l Bonnet-Guerrini\inst{1,2}
        \and Bruno Sanchez\inst{2}
        \and Dominique Fouchez\inst{2}
        \and Benjamin Racine\inst{2}
        \and Maya Guy\inst{3,4}
        \and Mariam Sabalbal\inst{5}
        \and Manal Yassine\inst{2}
        \and Vincenzo Piuri\inst{1}
        }

   \institute{Universit\`a degli Studi di Milano, Department of Computer Science, Milan 20133, Italy\\
             \email{raphael.bonnet-guerrini@unimi.it}
             \thanks{Corresponding author}
            \and Aix Marseille Univ, CNRS/IN2P3, CPPM, Marseille, France
            \and Université Côte d’Azur, INRIA, CNRS, Laboratoire J.A.Dieudonné, Maasai team, Nice 06000, France
            \and Université Côte d’Azur, Observatoire de la Côte d’Azur, CNRS, Laboratoire Lagrange, Nice 06000 France
            \and Université de Liège, STAR Institute, Liège 4000, Belgium
            \\ }

  \date{}

  \abstract
    {Time-domain surveys generate large numbers of transient candidates, making Real-Bogus classification a critical step in automated discovery pipelines. Obtaining reliable human-labeled training sets is costly, while community-provided labels can be noisy and survey-dependent.}
   {We aim to develop a Real-Bogus classification framework that can be trained without human-labeled data using physically motivated injected transients and bogus-dominated survey data, that remains robust under strong class contamination, and that provides well-calibrated and reliable uncertainty quantification suitable for downstream decision-making.}
   {We combine simulated transient injections with a heavily contaminated survey class and train a dual-network model using an asymmetric co-teaching strategy for classes with different label-noise levels. We evaluate performance on a labeled benchmark subset and analyze the learned representation with latent-space visualization tools. For uncertainty quantification (UQ), we compare standard approaches (MC dropout and deep ensembles) and propose a low-cost hybrid strategy that exploits the dual-network setting to improve calibration. We further extend the evaluation to the light-curve domain to assess recovery of light-curve classes.}
    {The method achieves strong Real-Bogus performance on the labeled subset and remains stable under severe class contamination. It recovers transient light-curve classes with high fidelity, while single-source identification is fundamentally limited by intrinsic ambiguity in light-curve-derived labels. Our hybrid UQ approach achieves competitive calibration relative to substantially more expensive ensemble baselines. Latent-space analyses indicate that uncertainty aligns with the decision boundary and reveal that the model is able to identify subclasses within the bogus population.}
    {Our results show that injection-driven, weakly supervised training can enable scalable and consistent Real-Bogus classification without human-labeled training data while providing calibrated uncertainties. The methodology is well suited for transfer to forthcoming surveys by re-running the injection-based training pipeline.}

\maketitle

\section{Introduction}
\label{sec:intro}

\begin{figure}[h]
    \centering
    \includegraphics[width=1\linewidth]{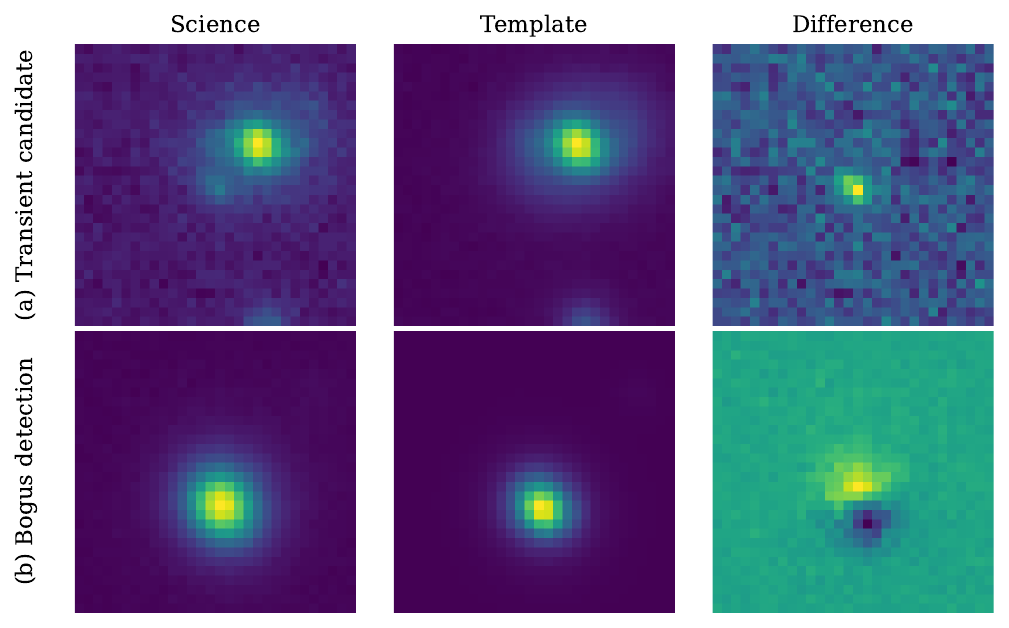}
    \caption{Example of difference imaging analysis (DIA) stamps. Each triplet shows the \textbf{Science} image (left), the \textbf{Template} image (middle), and the \textbf{Difference} \emph{(Science$-$Template$=$Difference)} (right). Top: a real transient produces a compact, PSF-like positive residual in the subtraction. Bottom: a bogus detection exhibiting a structured residual (dipole-like), characteristic of subtraction artifacts.}
    \label{fig:DIA}
\end{figure}

The new Vera C. Rubin Legacy Survey of Space and Time (LSST) \citep{techreport_LSST_2009, Ivezic2019:LSST} features a time-domain component that will detect transients throughout its 10-year run using Difference Image Analysis (DIA) \cite{lsst_science_book_2009, Ivezic2019:LSST, DIA_lsst_2024}. In DIA, a template image is subtracted from a newly observed image to identify changes associated with new or variable sources \citep{DIA_Alard_1998, DIA_alard_2000}. This method is crucial for detecting transient events in crowded fields or under varying observational conditions. Typically, supernovae occupy the short-duration region of the luminosity-timescale phase space and therefore benefit from high-cadence observations that can identify them before they fade \citep{2009transientslocaluniverse}. 
Because DIA depends on accurate image alignment and point-spread-function matching, it also produces false positives ("bogus") arising from noise, artifacts, imperfect subtraction, cosmic rays, bad pixels, and atmospheric effects. Illustrated examples of spurious and successful DIA are shown in Fig.~\ref{fig:DIA}.
 
In this work, we use the term source to denote a single-epoch DIA detection, and the term object to denote the association of sources across epochs at a consistent sky position; the magnitudes of the sources associated with a given object across time form a light curve.

 While light curves provide the richest information for transient typing, they are typically only informative after multiple detections have been accumulated. Relying solely on light-curve-based methods typically implies that identification occurs only after, or at best late in, the visible window of the transient. However, many downstream science goals require rapid spectroscopic follow-up \citep{spectro_lsst_2013, science_lsst_2016}. 
 At LSST-scale, alert brokers and downstream transient pipelines require early transient-bogus discrimination to suppress spurious detections and prioritize candidates before applying more computationally intensive light-curve classifiers \citep{ANTARES_2018, Fink_2020, Alerce_Carrasco_Davis_2021, broker_overview_2025}.

Historically, distinguishing real transients from bogus detections has relied on a combination of algorithmic quality flags and manual human inspection. With the rate of detection expected from the Vera C. Rubin Observatory, manual inspection is not feasible \citep{key_number_lsst_2024}.

In recent years, machine learning has demonstrated remarkable success in various domains, including image classification. Traditional image classification relied heavily on hand-crafted features and domain expertise. However, with the advent of Deep Learning (DL), convolutional neural networks (CNNs) \citep{lecun1989CNN} have become the standard for image classification tasks \citep{Hinton2012:ImageNet}, including astronomical photometry applications \citep{Dieleman_2015_cnn_gal_morpho, CNN_lensing_lanusse_2018}. CNNs automatically learn hierarchical features from raw pixel data, significantly improving accuracy and reducing the need for manual feature engineering.
Following these advances, DL methods have been developed for real-bogus classification \citep{CNN_TRANSIENTS_2018, LRP_bogustransient_2018, Alerce_Carrasco_Davis_2021, meercrab_transients_2021, BCNN_2021, CNN_DT_bogus_transient_2022, CNN_TRANSIENTS_2022, transient_vit_cnn2023}.

Despite these advancements, we identify \textbf{two main challenges} for deploying DL-based methods for real-bogus classification.\vspace{-2ex}

\paragraph{Challenge (i)} DL often requires \textbf{large labeled training sets}, which can be expensive and time-consuming to obtain. Real-bogus classification typically suffers from lack of labeled data, most methods in the literature therefore rely on supervised learning with human-labeled datasets \citep{CNN_TRANSIENTS_2018,LRP_bogustransient_2018,Alerce_Carrasco_Davis_2021,meercrab_transients_2021,CNN_DT_bogus_transient_2022,CNN_TRANSIENTS_2022, transient_vit_cnn2023}. 
While transient simulation through artificial source injection is well established \citep{injection_DES_Kessler_2015, LSST_simulation_Sanchez_2022}, bogus artifacts are heterogeneous, survey-specific, and difficult to model realistically. In addition, cross-survey domain shift has been observed for real-bogus classification, implying that methods trained for one telescope do not necessarily transfer directly to another \citep{cabreravives2023domainadaptationminimaxentropy}.
To meet the labeling demands of DL in the large-survey era, astronomy has increasingly relied on collaborative labeling. While successful, these efforts still incur significant overhead and produce labels with non-negligible inter-annotator disagreement, requiring careful aggregation and bias correction \citep{galaxyzoo_Lintott_2010}.

The use of simulated data has been explored in \cite{BCNN_2021} for point spread functions (PSFs) of minor planets superimposed on galaxy images.
To date, unsupervised approaches have been tested \citep{DESOM_transient_2022}, as have active-learning methods that reduce the amount of labeled data required to achieve strong performance \citep{active_learning_transient_2025}.
Nevertheless, real-bogus classification remains largely constrained by the need for survey-specific human-labeled training data.

To overcome \textit{challenge (i)}, we propose a \textbf{human-label-free} approach based on the following intuition: the unlabeled survey detections are dominated by bogus examples, while realistic transient examples can be generated through source injection.
Transient appearances can be simulated realistically at the image level through injections.
By treating survey detections as a noisy bogus class and injected examples as transients during training, we formulate the task as a \textbf{weakly supervised learning (WSL)} problem with class-dependent label noise.
Building on existing WSL methods, we introduce \textbf{Asym-Co-teaching}, a variant of co-teaching designed for asymmetric, class-dependent noise. To evaluate the method, we construct training sets with different contamination levels and compare the performance of competing approaches across these settings.

\paragraph{Challenge (ii)} DL models are often treated as \textbf{black boxes} because their decisions arise from many nonlinear transformations over high-dimensional internal representations \citep{Black_box_Lipton}. Most explainable AI (XAI) methods explain a prediction by using a simplified proxy (or attribution) mechanism around a specific input \citep{ribeiro2016whyitrustyou, shrikumar2017justblackboxlearning, binder2016layerwiserelevancepropagationneural, lundberg2017unifiedapproachinterpretingmodel}. Such techniques have been explored for real-bogus transient classification \citep{LRP_bogustransient_2018}, but they are inherently local: they provide insight for a single image and struggle to explain the global behavior of the model.

Recently, mechanistic interpretability has emerged as an approach for identifying global structures in the computations learned by neural networks \citep{MI_Transformer, circuit_thread_2020, sharkey2025openproblemsmechanisticinterpretability}. Early work often attempted to associate individual neurons, filters, or layers with interpretable behaviors \citep{CNN_layer_MI_2013, feature_pattern_visu_2017}, but the recognition of polysemanticity \citep{2022toymodelssuperposition} has increasingly shifted attention toward the geometry of latent representations. In this paper, we use latent space to denote the internal feature representation learned by an intermediate layer of the model.

Beyond interpretability, standard deterministic \textbf{deep learning models usually return point estimates} and do not, by default, provide well-calibrated predictive uncertainty. Yet scientific inference requires uncertainty quantification (UQ) that can be propagated to downstream analyses. As highlighted in  \citep{DESC_AI_2026}, adapting UQ methods to astronomy-specific settings is therefore essential for trustworthy scientific analyses. This is especially critical for cosmology-oriented transients targeted in this study (e.g., Type~Ia supernovae and other rare events used for population-level inference), where real-bogus classification occurs early in the pipeline: missed detections and, more importantly, systematic biases in detection or classification efficiency as a function of brightness, host-galaxy properties, or redshift directly affect the survey selection function and can propagate to bias cosmological measurements.

To address \textit{challenge (ii)}, we probe the model through \textbf{latent-space analysis} using dimensionality-reduction and visualization tools to seek global structure in the learned representation. Because interpretability analyses are primarily qualitative, we complement them with a systematic evaluation of \textbf{uncertainty quantification (UQ)} methods. This comparison is motivated by two considerations: first, epistemic uncertainty provides a natural representation of lack of knowledge in scientific inference, especially when ground truth is incomplete; second, uncertainty estimates in deep learning are often fragile, potentially miscalibrated, and sensitive to distribution shift. We therefore assess their reliability empirically in our setting and adapt existing UQ methods to the specific structure of Asym-Co-teaching.

In this paper, we address these two challenges as follows. Section~\ref{sec:data-inj} introduces the dataset, the injection procedure, and the construction of the ground truth. Section~\ref{sec:DL-classification} then presents the Real-Bogus classification task, the model architecture, and the training and optimization strategy. Next, we describe in Section~\ref{sec:wsl} our experimental methodology for evaluating WSL approaches including the Co-teaching framework and Asym-Co-teaching, our adaptation of Co-teaching to class-dependent asymmetric noise. In Section~\ref{sec:UQ}, we propose a methodology to assess UQ in this setting and adapt existing UQ methods to Asym-Co-teaching. We present the latent space exploration and visualization in Section~\ref{sec:latent}. The results are presented in Section~\ref{sec:results}, including UQ and WSL experiments as well as an extension to light-curve classification and inference. We discuss the implications and limitations of our findings in Section~\ref{sec:discussion}, and conclude in Section~\ref{sec:conclusion}.

\section{Data and injection process}    
\label{sec:data-inj}
We use imaging data from the Hyper Suprime-Cam (HSC) on the 8.2\,m Subaru Telescope, located on Maunakea, Hawai`i \citep{HSC_2018}. HSC comprises 104 science CCDs of size $2048\times4096$ pixels with a pixel scale of $0.168~\mathrm{arcsec\,pix^{-1}}$ \citep{HSC_2018}. For this study, we use two HSC-based datasets processed with the LSST Science Pipelines \citep{LSST_simulation_Sanchez_2022, DIA_lsst_2024}, whose DIA implementation follows the image-subtraction framework of \citet{DIA_Alard_1998}. To keep end-to-end processing computationally tractable during method development, we use \textit{rc2\_subset} as the main training dataset. This is a subset of the HSC SSP survey, used for regular testing of the LSST Data Release Production, and consists of six central science CCDs in the COSMOS field for eight visits in each of the $grizy$ filters (40 visits in total). For inference and light-curve follow-up, we use a larger HSC-UDEEP COSMOS subset containing 103 science CCDs over 862 visits. All images are processed through the LSST Science Pipelines, including difference image analysis (DIA), source detection, and measurement.

\subsection{Difference Image Analysis}

The first step of difference-image analysis (DIA) is to construct a deep template (reference) image by coadding multiple exposures of the same field; the template represents the static sky for that field. A science image (single-epoch exposure) is a new observation of the same field acquired at a later epoch. The science image is aligned photometrically and astrometrically to the template, after which PSF matching and subtraction are performed. Ideally, the resulting difference image contains only sources that have changed between the two epochs (Fig.~\ref{fig:DIA}).

In practice, most detections on difference images are spurious. Common sources of bogus detections include imperfect astrometric registration, PSF or photometric mismatches, variable atmospheric conditions, detector defects, cosmic rays, and subtraction artifacts. Robust Real-Bogus filtering is therefore essential to suppress these artifacts before downstream transient characterization.

From the resulting difference images, candidate point-like detections are identified and represented as $30\times30$ pixel cutouts.
Each cutout is normalized independently by clipping pixel values to the 1st-99th percentile range, applying an $\operatorname{arcsinh}$ transform, and then performing robust standardization using the cutout median and median absolute deviation. These cutouts form the input to our real-bogus classifier. 

\subsection{Supernova-Like injection}
\label{sec:sn_inj}

Many injection pipelines generate source catalogs by sampling positions and magnitudes from generic or ad hoc distributions. In this work, we instead adopt an injection scheme tailored to supernova-like point sources. This choice reflects the scope of our analysis: our primary downstream application is the detection of SNe~Ia-like transients, and we therefore design the injection population to approximate the region of astrophysical parameter space most relevant to that use case. The resulting injection model combines host-galaxy information with simple priors on source location and brightness.

Host galaxies are identified using the LSST pipeline \textit{extendedness} criterion, a binary flag (0 = point-like, 1 = extended) based on a threshold on the ratio between PSF flux and model flux. For each selected galaxy we estimate the size and orientation of its projected elliptical profile, and use these quantities to define the injection-position prior.

The projected distance of the injection from the galaxy center is sampled from a normal distribution centered on the host position, with variance set by the semi-major axis of the galaxy's elliptical shape. Mathematically, the offset $d_{\text{inj}}$ is sampled as $ d_{\text{inj}} \sim \mathcal{N}(\mu = 0, \sigma^2 = a^2) $, where $ a $ is the semi-major axis of the galaxy. This produces a host-associated population of synthetic point sources whose projected offsets follow the size and orientation of the underlying galaxy. We then sample the injected magnitude as $ m_{\text{inj}} \sim \mathcal{U}(m_{\text{host}}-1,\; m_{\text{host}}+3) $, which is intended to span a broad range of source-to-host contrast ratios while remaining tied to the observed host population.

In addition, we include a 10\% hostless component. For these injections, positions are drawn uniformly over the image footprint, and magnitudes are sampled independently in each band from empirical distributions fitted to the UDEEP catalog. This component is intended to prevent the training set from being restricted to host-associated events only.

The resulting injection catalogs are ingested by the LSST pipeline middleware and injected at the image level as part of the standard processing workflow\footnote{\href{https://pipelines.lsst.io/modules/lsst.source.injection/index.html}{lsst.source.injection}}. The injected exposures are then passed through the same DIA, detection, and measurement steps as the real survey data.

To obtain an approximately balanced training set at the detection level, we tune the number of injected sources so that the number of recovered injected detections is comparable to the number of detections obtained from the corresponding real survey processing. In \textit{rc2\_subset}, the real survey processing contains $43\,607$ detections, while our injection procedure produces $N_{\rm inj}=89\,570$ artificial point sources, of which $51\,550$ are recovered by the pipeline. The unrecovered injections predominantly populate the faint end of the injected distribution (Appendix~\ref{app:magnitude}), and the recovered sample remains close to balanced relative to the survey detections.

To verify that injections remain sparse at the image level, we estimate the affected pixel fraction. Averaging over the $40\times6=240$ CCD images gives $\langle N_{\rm src}\rangle \simeq 373.2$ injections per CCD. Approximating the footprint of each source by a disk of radius $R=3\,{\rm FWHM}\simeq12.5$ pixels affects a fractional area of about $2\%$. This is an approximate upper bound, since source footprints may overlap, and indicates that injections remain sparse on the scale of the full detector.

We combine the recovered injection-driven detections with detections from the corresponding real survey processing (i.e. detections not associated with injected truth) to form the \textbf{baseline training set}. The baseline training set therefore consists of single-detection cutouts with two training labels: recovered injections are labeled \textbf{Transient}, while all survey detections are labeled \textbf{Bogus}, although this latter class is expected to contain genuine transients that the model aims to recover at inference time.

\subsection{Constructing the evaluation set}
\label{sec:groundtruth}

Because our method relies on a noisy class in which genuine transients are mislabeled during training, we need an independent ground truth for evaluation. We therefore construct a manually curated evaluation set from HSC-UDEEP light curves. 
Neither human labels nor light curve level information is used for model training and inference, filtering and manual inspection is used only at the evaluation stage.
Since \textit{rc2\_subset} does not provide sufficiently sampled light curves for this purpose, we derive the evaluation set from the more densely sampled UDEEP dataset. We first filter the UDEEP light curves to isolate well-measured SN-like candidates, and then manually inspect the remaining objects to define the set of real detections used for evaluation.

\paragraph{Light-curve filtering.} 
We apply a sequence of cuts to isolate a compact set of well-sampled transient-like light curves for manual inspection. First, we remove individual source measurements with signal-to-noise ratio ${\rm SNR}<5$. Objects with no remaining measurements above this threshold are discarded.

Next, we identify the epoch of maximum flux for each object and retain only light curves whose informative measurements lie within a window of $[-30,+100]$ days around maximum light. We then require at least six distinct observing nights with valid detections and reject objects whose mean PSF flux across retained epochs is negative, as these are indicative of systematically spurious or over-subtracted signals. This cut may remove genuine astrophysical negative residuals or fading events, such as disappearing sources or long-term variables. These objects are outside the scope of the present high-purity SN-like curated set, which is designed to evaluate the recovery of positive transient excesses rather than to provide a complete census of variable phenomena.

\begin{table}[h]
\centering
\small
\setlength{\tabcolsep}{6pt}
\renewcommand{\arraystretch}{1.2}
\begin{tabular}{l|cc}
\hline
\textbf{Step} & Objects discarded & Objects remaining\\
\hline
\textbf{$\text{SNR}<5$    }       & $157,112$ & $3,699,427$ \\
\textbf{Time window}              & $568,540$ & $3,130,887$ \\
\textbf{6 Minimum nights}         & $3,130,064$\tablefootnote{The large number of sources discarded at this step is mostly due to single observations, which account for $2,959,860$ sources in UDEEP.} & $823$ \\
\textbf{Negative flux}            & $352$ & $471$ \\
\textbf{Point source host}        & $95$ & $376$ \\
\textbf{Low flux ratio}          & $34$ & $342$ \\
\hline
\end{tabular}
\caption{Summary of the filtering steps used for light-curve selection for human labeling. For each step, we report the number of objects discarded and the number remaining afterward.}
\label{tab:lc-filtering}
\end{table}
To further suppress contamination, we incorporate host-galaxy information from the HSC deep coadds. For each object, we identify the epoch of maximum absolute difference flux and match its sky position to the nearest template source within $1^{\prime\prime}$. Objects are retained if they are either hostless or associated with an extended host, and if the transient-to-host flux ratio exceeds 1.4 in the $i$ band. These cuts intentionally favor well-measured SN-like transients over low-contrast or ambiguous cases and therefore define a high-purity evaluation subset rather than a representative sample of the full alert stream.
The numbers of discarded objects are reported in Table~\ref{tab:lc-filtering}.

\begin{table}[h!]
\centering
\small
\setlength{\tabcolsep}{4pt}
\renewcommand{\arraystretch}{1.1}
\begin{tabular}{@{}lcc@{}}
\toprule
Class & Objects $(N=306)$ & Sources $(N=4{,}820)$  \\
\midrule
SN-like            & $33$ $(10.8\%)$   & $1{,}211$ $(25.1\%)$ \\
OT                 & $41$ $(13.4\%)$   & $1{,}340$ $(27.8\%)$ \\
\textbf{Transient} & $74$ $(24.2\%)$   & $2{,}551$ $(\textbf{52.9\%})$ \\
\midrule
\textbf{Bogus}     & $232$ $(75.8\%)$  & $2{,}269$ $(\textbf{47.1\%})$ \\
\midrule
Total              & $306$ $(100\%)$   & $4{,}820$ $(100\%)$ \\
\bottomrule
\end{tabular}
\caption{Detailed class composition of the manually curated evaluation set after manual labeling and removal of objects labeled as Unknown. OT denotes other transient or variable objects. We report both the number of objects and the number of source detections in each class.}
\label{tab:groundtruth-set}

\end{table}

\paragraph{Human labeling.}

The 342 filtered light curves are then inspected manually using the coadded template image, band-averaged science and difference images, and the multi-band light curve itself (See App.~\ref{app:lc-tool}). We assign each object to one of four categories: \textbf{SN-like}, \textbf{Other transient or variable}, \textbf{Bogus}, or \textbf{Unknown}. Objects labeled \textbf{Unknown} are excluded from the binary evaluation set, leaving 306 objects and 4,820 source detections for quantitative analysis. 

\begin{figure*}[!t]
  \centering
  \resizebox{\textwidth}{!}{\subimport{texplot/}{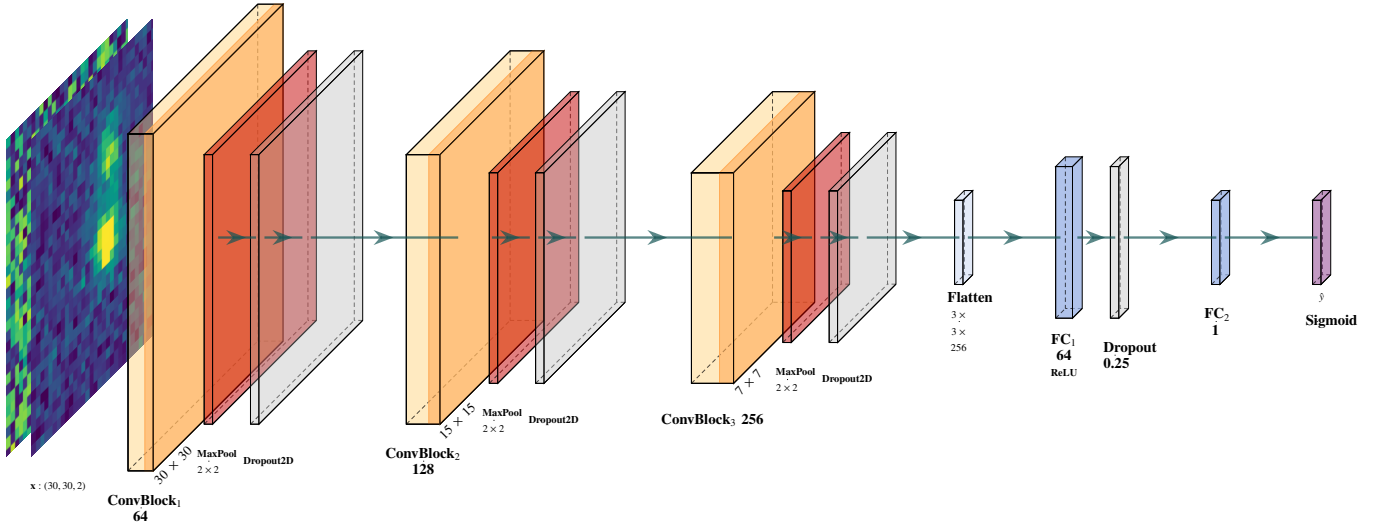}}%

  \caption{Schematic of the CNN architecture selected by hyperparameter optimization.}  
  \label{fig:CNN-architecture}
\end{figure*}

Although 232 of the 306 objects are labeled \textbf{Bogus}, the evaluation set is nearly balanced at the source-detection level because transient light curves typically produce detections over more epochs. Bogus-labeled objects can also contribute multiple detections, for example when subtraction residuals at a fixed location near a bright star recur across epochs owing to imperfect masking, PSF matching, or astrometric registration. These residuals often appear as dipole-like artifacts rather than isolated single-epoch events. We emphasize that this set serves as a manually curated evaluation set rather than spectroscopic truth.


\section{Deep Learning classification}
\label{sec:DL-classification}

\subsection{Classes and Confusion Matrix}

After injection, the training data consist of two labeled groups:
\begin{itemize}
  \item injected detections, labeled as \textbf{Transient};
  \item survey detections, labeled as \textbf{Bogus}, although this class is contaminated by genuine transients present in the survey data.
\end{itemize}

The objective of training is therefore not simply to separate injected from survey detections, but to learn a decision rule that separates transient-like from bogus-like detections despite label noise in the survey class. 

At training time, the confusion matrix must be interpreted relative to the observed training labels, not as a ground-truth confusion matrix:
\begin{itemize}
  \item True positives (TP) are injected data classified as transient.
  \item True negatives (TN) are survey data classified as bogus. 
  \item False positives (FP) are survey data classified as transient. 
  \item False negatives (FN) are injected data classified as bogus. 
\end{itemize}

In this setting, the FP category is of particular interest, because it contains the survey detections reclassified by the model as transient and may therefore include genuine transients hidden inside the noisy bogus-labeled class. By contrast, the FN category corresponds to injected detections that the model fails to recover as transients. Since the injected class is treated as the clean reference class during training, understanding and minimizing this failure mode is important.


\subsection{Network Architecture}
\label{sec:cnn-arc}

The network architecture consists of three convolutional blocks ($L = 3$). Each block comprises a $3\times 3$ convolutional layer, followed by batch normalization, a ReLU activation, $2\times 2$ max pooling, and spatial dropout regularization. The number of filters in the three convolutional layers is parameterized by a single base width $F$, and follows a geometric progression $(F \rightarrow 2F \rightarrow 4F)$. This coupling enforces a monotonic increase in channel width and avoids degenerate configurations while still allowing the overall capacity of the model to be controlled through $F$.

Each input $\mathbf{x} \in \mathbb{R}^{30 \times 30 \times 2}$ is a two-channel cutout, constructed by stacking the difference and template images (coadded). We do not include the science image explicitly, in order to reduce computational cost and input dimensionality while preserving both the subtraction residual and the static host-galaxy context.

The convolutional feature extractor computes
\begin{equation*}
\mathbf{h}_L
= {\mathrm{ConvBlock}_3 \circ \mathrm{ConvBlock}_2 \circ \mathrm{ConvBlock}_1}(\mathbf{x}),
\end{equation*}

where each $\mathrm{ConvBlock}_\ell$ consists of convolution, batch normalization, ReLU, max pooling, and dropout. After convolutional feature extraction, the feature maps are flattened and passed through two fully connected layers, where the hidden dimensionality depends on the spatial reduction factor and a configurable hidden size, as will be discussed next. The classifier computes
\begin{equation*}
\begin{aligned}
f(\mathbf{x}) &= \mathrm{FC}_2\;\!\Big(
  \mathrm{Dropout}\Big(
    \mathrm{ReLU}\Big(
      \mathrm{FC}_1\big(\mathrm{Flatten}(\mathbf{h}_L)\big)
    \Big)
  \Big)
\Big),
\end{aligned}
\end{equation*}
and the final prediction is obtained via
\begin{equation*}
\hat{y} = \sigma\!\big(f(\mathbf{x})\big),
\end{equation*}
with $\sigma(\cdot)$ denoting the sigmoid activation function for binary classification (transient vs. bogus detection).

\subsection{Hyperparameter Optimization}
\label{sec:hyperopt}

The CNN hyperparameters are tuned via Bayesian optimization using the Optuna framework \citep{optuna2019}, with the objective of minimizing the loss. To keep the search statistically efficient under a limited budget of 50 trials, we adopt a structured and low-dimensional hyperparameterization. Architecturally, only the base convolutional width $F$ (which determines the filter configuration $(F,2F,4F)$) and the number of units in the dense layer are tuned. Similarly, a single base dropout rate ($B$) controls regularization throughout the network: convolutional blocks apply scaled rates ($(0.5B, 0.5B, 0.5B)$), while the dense layer uses $B$. This geometric progression enforces smooth capacity scaling and reduces redundant architectural degrees of freedom.

The search is performed using Optuna's Tree-structured Parzen Estimator (TPE) sampler, introduced in \citep{TPE_2011}, which models the objective function with non-parametric density estimators in hyperparameter space. To reduce computational cost, the Bayesian optimization is carried out on a randomly selected $30\%$ subset of the available dataset, reserved specifically for hyperparameter search. 
The best configuration found by the search is summarized in Table~\ref{tab:best_hparams} and represented in Fig.~\ref{fig:CNN-architecture}. This configuration is used to train all the models we compare.
\begin{table}[h]
\centering
\begin{tabular}{l c}
\hline
Hyperparameter & Value \\
\hline
Batch size ($\texttt{batch\_size}$) & 128 \\
Learning rate ($\texttt{learning\_rate}$) & $1.62 \times 10^{-4}$ \\
Base filters $F$ ($\texttt{model\_params.base\_filters}$) & 64 \\
Dense units ($\texttt{model\_params.units}$) & 64 \\
Dense dropout ($\texttt{model\_params.base\_dropout}$) & 0.25 \\
\hline
\end{tabular}
\caption{Best architectural hyperparameters obtained from the Bayesian optimization.
\label{tab:best_hparams}}
\end{table}

\subsection{Early Stopping and Model selection strategy}

Early stopping is commonly based on validation loss or validation accuracy. In our setting, however, these criteria are not ideal because the survey-labeled bogus class is contaminated by genuine transients.
For model selection, we instead monitor the false negative rate (FNR) on the injected class, that is, the fraction of injected detections classified as bogus. This quantity is not used as the training objective itself; rather, it is used as an early-stopping and checkpoint-selection criterion. The intuition is that the injected class has the most trustworthy labels available during training, so preserving high recall on this class helps avoid selecting models that overfit the noisy survey labels.
We therefore use injected-class FNR as a proxy criterion for checkpoint selection, while the network itself is still optimized using the standard training loss. This strategy biases model selection toward sensitivity to transient-like detections without explicitly rewarding memorization of the noisy bogus-labeled survey class.

%
%
\section{Weakly Supervised Learning for Injection-based Transient--Bogus Classification}
\label{sec:wsl}
\begin{table*}[ht]
\centering
\begin{tabular}{lccccccc}
\hline\hline
& & & \multicolumn{2}{c}{Observed training labels} & \multicolumn{2}{c}{True provenance} & \\
Dataset & $\eta$ [\%] & Total $N$
& $N(\tilde{y}=1)$ & $N(\tilde{y}=0)$
& $N(y=1)$ & $N(y=0)$ & $N_{\mathrm{mis}}$ \\
& & & (inj.-labeled) & (survey-labeled)
& (injection-origin) & (survey-origin) & \\
\hline
Baseline      & 0.00  & $95\,157$ & $51\,550$ & $43\,607$ & $51\,550$ & $43\,607$ & $0$ \\
Low-noise     & 15.09 & $68\,821$ & $34\,398$ & $34\,423$ & $39\,593$ & $29\,228$ & $5\,195$ \\
Medium-noise  & 24.99 & $44\,666$ & $22\,349$ & $22\,317$ & $27\,927$ & $16\,739$ & $5\,578$ \\
High-noise    & 35.23 & $30\,475$ & $15\,262$ & $15\,213$ & $20\,621$ & $9\,854$  & $5\,359$ \\
\hline
\end{tabular}
\caption{
Summary of the datasets used to evaluate robustness to controlled asymmetric label noise.
For each dataset, we report both the observed training labels $\tilde{y}$ (used for optimization) and the true provenance labels $y$ (used only for analysis). 
Noise is introduced by flipping a subset of injection-origin samples from $\tilde{y}=1$ to $\tilde{y}=0$, so that the survey-labeled class contains a controlled fraction of hidden injection-origin samples.
The contamination fraction is
$\eta = P(y=1 \mid \tilde{y}=0) = N_{\mathrm{mis}}/N(\tilde{y}=0)$,
where $N_{\mathrm{mis}} = N(y=1,\tilde{y}=0)$ is the number of injection-origin samples relabeled as survey-labeled for training.
The total number of samples decreases with increasing noise because survey-origin samples are subsampled to keep the two training-label classes approximately balanced.
}
\label{tab:noise_datasets}
\end{table*}
Supervised classifiers can be sensitive to label noise. While moderate levels of corruption may be tolerated \citep{noiseDL_2017}, they can bias the learned decision boundary and degrade performance \citep{noise_gen_2017}.
As calibration pipelines improve in modern wide-field surveys \citep[e.g.,][]{SDSS_calib_2008, DES_calib_2017,Photo_cal_2024} and DIA implementations continue to reduce subtraction artifacts and false positives \citep{DIA_lsst_2024}, the bogus fraction among DIA candidates may decrease, and the survey-labeled class will contain a larger fraction of genuine transients, weakening our injection-based label-purity assumptions. 
We therefore study the robustness of our framework under controlled one-sided label corruption by artificially contaminating the survey-labeled training class with injection-origin samples.

The \textbf{baseline dataset}, described in Section~\ref{sec:sn_inj}, contains $51\,550$ injection-origin detections and $43\,607$ survey-origin detections, for a total of $95\,157$ examples. In this baseline, the injection provenance and the training label coincide.

For each example, we retain two binary variables:
(i) a provenance label $y$ (stored as \texttt{spy\_injected}), indicating whether the cutout truly originates from an injection ($y=1$) or from survey data ($y=0$); and
(ii) a training label $\tilde{y}$ (stored as \texttt{is\_injection}), which is the label used for model training and may be intentionally corrupted.
In the baseline dataset, $y=\tilde{y}$ for all samples.

To construct noisy variants, we apply a one-sided, class-conditional corruption mechanism: we randomly select $N_{\mathrm{mis}}$ injection-origin examples ($y=1$) and flip their training label from $\tilde{y}=1$ to $\tilde{y}=0$. This contaminates the survey-data training class with transient-like samples. Because labels are flipped in only one direction, we additionally subsample survey-origin examples so that the two training-label classes remain approximately balanced, \(N(\tilde{y}=1)\simeq N(\tilde{y}=0)\).

Using this procedure, we generate three datasets with $\eta \simeq 15\%$, $25\%$, and $35\%$ contamination in the survey-data class.  Table~\ref{tab:noise_datasets} summarizes both the training-label distribution ($\tilde{y}$) and the underlying provenance distribution ($y$).


\subsection{Co-teaching}

Co-teaching~\citep{coteaching_han2018} is a method used in machine learning to handle datasets with noisy labels. Two models are initialized and trained simultaneously on the same dataset. During each training iteration, each model selects a subset of training samples with the smallest loss (i.e., the samples it is most confident about). Each network then uses the small-loss samples selected by the other network to update its parameters.

The intuition is that samples with smaller losses are more likely to be correctly labeled, while samples with higher losses are more likely to be noisy. This cross-teaching mechanism ensures that networks do not reinforce their own biases and are less likely to overfit to noisy labels. Over time, as the networks learn, they become better at identifying clean samples, and the training process becomes more robust to label noise\footnote{All parameter configurations for each Weakly Supervised Learning and Uncertainty Quantification method are presented in Appendix~\ref{app:param}.}.

In our setting, standard Co-teaching has two practical limitations. First, it requires the user to specify a forget rate, i.e. the fraction of samples assumed to be corrupted; in our experiments this quantity is set using prior knowledge from the controlled-noise setup (Appendix~\ref{app:param}). Second, the sample-selection step is performed globally within each mini-batch, without explicitly accounting for class-dependent noise asymmetry. In our application, however, the noise is concentrated primarily in the survey-labeled bogus class, while the injection-labeled transient class is intended to remain comparatively clean. As a result, the standard selection rule may discard informative transient examples rather than focusing rejection on the noisy class. A formal description of standard Co-teaching is given in Appendix~\ref{app:coteaching-form}.

\subsection{Asymmetric Co-teaching}
\label{sec:asym_coteaching}
In our application, the label noise is highly asymmetric: the fraction of corrupted labels is much higher in the \emph{bogus} class than in the \emph{transient} class. In addition, for our target use case, missing a true transient is typically more costly than passing some additional false detections to downstream filtering stages.

Standard Co-teaching uses a single forget rate and selects small-loss samples over the whole mini-batch, independently of their class. As a result, there is no guarantee that the high-loss samples discarded during training are predominantly noisy examples from the noisy majority class (bogus); the procedure may instead discard rare but informative transient examples that happen to incur higher loss. 
To address this issue, we propose \textbf{Asymmetric Co-teaching} (Asym-Co-teaching), which extends Co-teaching by introducing class-specific forget rates and class-wise selection of small-loss samples.
\begin{figure}[h!]
    \centering
    \resizebox{1\linewidth}{!}{\subimport{texplot/}{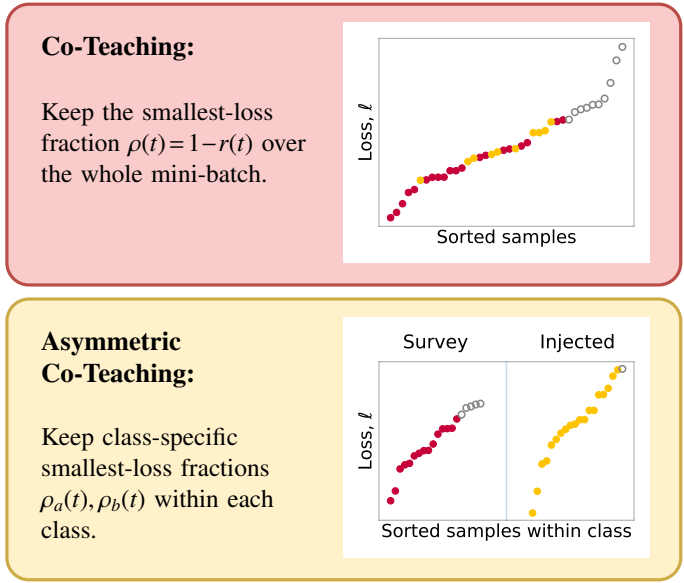}}
    \caption{Illustration of sample-selection rules in Co-teaching variants. Top: standard Co-teaching applies a class-independent forget rate and selects small-loss samples globally within the mini-batch. Bottom: Asym-Co-teaching performs class-wise small-loss selection with class-specific forget rates. Hollow markers denote discarded samples.}
    \label{fig:co-teaching}
\end{figure}

Let:
\begin{equation*}
\mathcal{B}_t = \{(x_i, \tilde{y}_i)\}_{i \in I_t}
\end{equation*}
denote the mini-batch at epoch \(t\), where \(I_t\) is the index set of samples in the batch and \(\tilde y_i \in \{a,b\}\) is the observed training label. We define the class-specific index sets
\begin{equation*}
I_{t,a} = \{ i \in I_t : \tilde y_i = a \},
\qquad
I_{t,b} = \{ i \in I_t : \tilde y_i = b \}.
\end{equation*}
The per-sample losses of network~1 and network~2 on this batch are defined by their Binary Cross Entropy ($\mathbf{BCE}$) as
\begin{equation*}
\ell_i^{(1)}(t) = \mathbf{BCE}\big(f_{\theta_1}(x_i), \tilde y_i\big),
\qquad
\ell_i^{(2)}(t) = \mathbf{BCE}\big(f_{\theta_2}(x_i), \tilde y_i\big),
\qquad i \in I_t.
\end{equation*}

We introduce class-specific forget rates
\begin{equation}
\begin{aligned}
r_a(t) &= \min\!\left(r_a \cdot \frac{t}{T_k},\, r_a\right),
\\
r_b(t) &= \min\!\left(r_b \cdot \frac{t}{T_k},\, r_b\right),
\end{aligned}
\end{equation}
Similarly to Co-teaching, the fractions of samples kept in each class at iteration $t$ are therefore $1 - r_a(t)$ and $1 - r_b(t)$, and we set
\begin{equation*}
k_{t,a} = \big\lfloor (1 - r_a(t))\, |I_{t,a}| \big\rfloor,
\qquad
k_{t,b} = \big\lfloor (1 - r_b(t))\, |I_{t,b}| \big\rfloor.
\end{equation*}

For network~1, we sort the samples within each class according to their loss:
\begin{equation*}
\ell_{i_{1,a}^t}^{(1)}(t) \le \ell_{i_{2,a}^t}^{(1)}(t) \le \cdots \le \ell_{i_{|I_{t,a}|,a}^t}^{(1)}(t),
\end{equation*}
\begin{equation*}
\ell_{i_{1,b}^t}^{(1)}(t) \le \ell_{i_{2,b}^t}^{(1)}(t) \le \cdots \le \ell_{i_{|I_{t,b}|,b}^t}^{(1)}(t),
\end{equation*}
where $(i_{1,a}^t,\dots,i_{|I_{t,a}|,a}^t)$ is a permutation of $I_{t,a}$ and $(i_{1,b}^t,\dots,i_{|I_{t,b}|,b}^t)$ is a permutation of $I_{t,b}$. The sets of reliable samples selected by network~1 at iteration $t$ are then
\begin{equation*}
\mathcal{R}_{1,t,a} = \{ i_{1,a}^t, \dots, i_{k_{t,a},a}^t \},
\qquad
\mathcal{R}_{1,t,b} = \{ i_{1,b}^t, \dots, i_{k_{t,b},b}^t \}.
\end{equation*}
Equivalently, they can be characterized as
\begin{equation*}
\mathcal{R}_{1,t,a}
= \arg\min_{\substack{S \subset I_{t,a} \\ |S| = k_{t,a}}}
    \sum_{i \in S} \ell_i^{(1)}(t),
\qquad
\mathcal{R}_{1,t,b}
= \arg\min_{\substack{S \subset I_{t,b} \\ |S| = k_{t,b}}}
    \sum_{i \in S} \ell_i^{(1)}(t).
\end{equation*}
Analogously, using the losses $\ell_i^{(2)}(t)$ of network~2, we define the sets
$\mathcal{R}_{2,t,a}, \mathcal{R}_{2,t,b}$.

For each mini-batch, both networks compute individual losses, rank the samples by loss \emph{within each class}, and then train on the small-loss samples selected by the peer network. The Asymmetric Co-teaching losses are thus given by
\begin{equation}
\begin{aligned}
\mathcal{L}_{\text{Asym-co}}^{(1)}(t) 
&= \frac{1}{|\mathcal{R}_{2,t,a}| + |\mathcal{R}_{2,t,b}|}
\Bigg(
  \sum_{i \in \mathcal{R}_{2,t,a}} 
    \mathbf{BCE}\big(f_{\theta_1}(x_i), \tilde y_i\big)
\\
&\quad +
  \sum_{i \in \mathcal{R}_{2,t,b}} 
    \mathbf{BCE}\big(f_{\theta_1}(x_i), \tilde y_i\big)
\Bigg),
\\[0.4em]
\mathcal{L}_{\text{Asym-co}}^{(2)}(t) 
&= \frac{1}{|\mathcal{R}_{1,t,a}| + |\mathcal{R}_{1,t,b}|}
\Bigg(
  \sum_{i \in \mathcal{R}_{1,t,a}} 
    \mathbf{BCE}\big(f_{\theta_2}(x_i),\tilde y_i\big)
\\
&\quad +
  \sum_{i \in \mathcal{R}_{1,t,b}} 
    \mathbf{BCE}\big(f_{\theta_2}(x_i),\tilde y_i\big)
\Bigg).
\end{aligned}
\end{equation}

These class-specific forget rates determine the fraction of samples discarded from each class at epoch \(t\), with remember rates \(\rho_c(t)=1-r_c(t)\) controlling the number of samples retained for training.



%
\section{Uncertainty Quantification for injection-based Real-Bogus Classification}
\label{sec:UQ}

Uncertainty quantification is fundamental to scientific inference. Although machine learning can be viewed as an extension of statistical modeling, obtaining uncertainty estimates that are simultaneously well calibrated, predictive, and computationally efficient remains challenging, particularly in deep learning \citep{calibration_DL_guo_2017}. In this work, we evaluate and compare several uncertainty quantification approaches for injection-based real-bogus classification.

\subsection{Evaluation of the uncertainty methods}

We evaluate uncertainty methods along two complementary axes:  
(i) the quality of their probabilistic predictions, assessed with standard calibration and scoring metrics; and  
(ii) the extent to which the resulting uncertainty quantification correlates with physically meaningful indicators of classification difficulty.

Calibration metrics assess whether predicted probabilities are statistically reliable estimates of true class frequencies \citep{calibration_2005}.

For probabilistic evaluation, we report the negative log-likelihood (NLL), the Brier score, and the expected calibration error (ECE). NLL and Brier score are proper scoring rules, and therefore reflect both calibration and sharpness of probabilistic predictions, while ECE more directly quantifies the mismatch between predicted confidence and empirical accuracy.

For binary classification, the negative log-likelihood is
\begin{equation}
    \mathrm{NLL}=-\frac{1}{N}\sum_{n=1}^{N} \left[y_n \log \bar{p}_n+(1-y_n)\log(1-\bar{p}_n)\right],
\end{equation}
where \(\bar{p}_n\) denotes the predictive probability assigned to the positive class for sample \(n\), and \(y_n \in \{0,1\}\) is the observed label. Lower NLL indicates better probabilistic predictions and strongly penalizes overconfident errors, making it particularly sensitive to miscalibration.

The Brier score is
\begin{equation}
    \mathrm{BS}=\frac{1}{N}\sum_{n=1}^{N}(\bar{p}_n-y_n)^2.
\end{equation}
Lower Brier scores indicate better probabilistic predictions.

The expected calibration error (ECE) measures how well a model's estimated probabilities match the true probabilities. This is done by calculating the weighted average error of the estimated probabilities.
\begin{equation}
    \mathrm{ECE}=\sum_{m=1}^{M}\frac{|B_m|}{N}\left|\mathrm{acc}(B_m)-\mathrm{conf}(B_m)\right|,
\end{equation}
where \(B_m\) is the set of predictions falling in bin \(m\), \(\mathrm{acc}(B_m)\) is the empirical accuracy in that bin, and
\begin{equation*}
\mathrm{conf}(B_m)=\frac{1}{|B_m|}\sum_{i\in B_m}\bar{p}_i.
\end{equation*}

In addition to calibration, we assess the quality of our uncertainty quantification by examining its correlation with two physically meaningful and interpretable quantities: the signal-to-noise ratio (SNR) and the distance from maximum brightness for supernova samples. 

At low SNR, the pixel-level evidence in a single-epoch difference-image cutout may be insufficient to reliably distinguish a genuine PSF-like astrophysical transient from a chance noise fluctuation with similar morphology, implying an effective information limit for transient-bogus classification.

Similarly, sources far from the peak luminosity of a supernova are expected to be more difficult for the model to classify, because supernovae observed during their rise or decline phases have less distinctive photometric signatures. Note that this metric can be computed only for sources that are part of light curves labeled as SN-like (see Section~\ref{sec:groundtruth}), excluding sources that are unlabeled or classified as other transient or variable or Bogus.

While these two quantities are correlated, it is important to note that SNR is not evenly distributed across the training classes, as shown in Appendix~\ref{app:SNR_app}, and is therefore expected to follow a similar tendency. Using two metrics with different class-dependent behavior makes the evaluation more robust. 
To determine the correlation between the uncertainty estimator method and the physical quantities, we use the Spearman rank correlation coefficient $\rho$ computed as:
\begin{equation}
    \rho = 1 - \frac{6 \sum_{i=1}^{n} d_i^2}{n(n^2 - 1)},
\end{equation}
where $d_i = \text{rank}(x_i) - \text{rank}(y_i)$ is the difference between the ranks of corresponding values, and $n$ is the number of samples.

\subsection{Deep Ensembles}
\label{sec:deep-ensembles}

Deep Ensembles (or ensembles) provide a simple and effective approach to predictive uncertainty in deep learning \citep{deep_ensemble_lakshminarayanan2017}. Unlike explicitly Bayesian neural-network methods \citep{BNNmackay1992}, they require no modification of the base architecture and often provide strong empirical uncertainty quantification. They also differ from methods such as MC dropout because independent initialization and training encourage exploration of different regions of the loss landscape \citep{fort2020deepensembleslosslandscape}.

For our application, we are independently training CNN models. Each model, denoted by $f_m(\cdot)$, shares the same architecture (see Section~\ref{sec:cnn-arc}) but is initialized with different random weights. All models are trained separately on the same dataset following the standard procedure described previously. At inference time, the ensemble prediction is obtained by averaging the individual model probabilities.
Beyond improving prediction stability, the ensemble also facilitates uncertainty quantification. 
We quantify this uncertainty from the dispersion of the ensemble predictions. In practice, we use the standard deviation of the member probabilities for a given input \(\mathbf{x}\), with larger dispersion indicating greater predictive disagreement. Further details are given in Appendix~\ref{app:uq-formalism}.

\subsection{Repulsive Ensembles}
Repulsive Ensembles \citep{rep_ensemble_dangelo_2021} extend standard deep ensembles by introducing an interaction term during training that encourages ensemble members to represent diverse functions while still fitting the data well. In practice, the repulsion is imposed in function space rather than directly in weight space, since different parameter settings can correspond to similar predictive functions. 
The motivation is to increase functional diversity across ensemble members, thereby improving uncertainty estimates relative to ensembles that remain concentrated around a narrow region of the loss landscape. At inference time, predictions are aggregated in the same way as for standard deep ensembles, and uncertainty is quantified from the dispersion across ensemble members.

\subsection{MC Dropout}

Monte Carlo Dropout (MC Dropout) \citep{MCdropout2016} is a widely used uncertainty quantification method in deep learning.  It can be interpreted as a particular BNN construction in which Bernoulli dropout masks induce an approximate distribution over the weights; running the model multiple times with dropout enabled at inference then amounts to Monte Carlo sampling from the resulting predictive posterior.
A practical advantage of the method is that it requires no architectural change beyond the use of dropout during training. The same trained network can then be sampled multiple times at inference by applying different dropout masks. Further formal details are given in Appendix~\ref{app:uq-formalism}.

\subsection{Uncertainty for Co-teaching Methods}
\label{sec:unc_asy_coteaching}

As discussed above, deep ensembles provide a strong baseline for uncertainty quantification by capturing variability across different modes of the loss landscape, and they can be further diversified through repulsive training. 
However, these models are computationally expensive: an ensemble of $N$ networks requires $N$ full training runs and $N$ forward passes per sample at inference time. 
In the Co-teaching setting, two networks are already trained from different random initializations and can therefore be viewed as a small coupled ensemble, although they are not independent in the same sense as a standard deep ensemble. 
Because an ensemble of two networks may provide only limited diversity, we augment the Co-teaching pair with MC Dropout at inference time. This allows multiple stochastic predictions from each of the two trained models and increases the number of predictive samples at low additional training cost.
The predictive probability now becomes: 
\begin{equation}
    \bar{p}(\mathbf{x}^*) = \frac{1}{NM} \sum_{n=1}^{N} \sum_{m=1}^{M} p_{n,m}(\mathbf{x}^*)
    = \frac{1}{NM} \sum_{n=1}^{N}\sum_{m=1}^{M} \sigma\big(f_{n,m}(\mathbf{x}^*; \hat\theta_{n};z_{n,m})\big),
\end{equation}
where $N=2$ is the number of Co-teaching networks, from different initializations, composing our pseudo-ensemble, $z_{n,m}$ the Dropout Mask and $M$ is the number of Dropout masks that will be applied to both models. 
Following the same logic, we can estimate the uncertainty from the second centered moment: 
\begin{equation}
\mathrm{Var}_{\text{Co-Ens-Dropout}}(\mathbf{x}^*) =
\frac{1}{NM} \sum_{n=1}^{N}\sum_{m=1}^{M}
\big(p_{n,m}(\mathbf{x}^*) - \bar{p}(\mathbf{x}^*)\big)^2.
\end{equation}

This construction can be viewed as a heuristic mixture approximation in which each trained Co-teaching network defines one component and MC Dropout provides local stochastic samples around that component.

In a Bayesian framework as a mixture approximate posterior, the $N$ networks converge to different solutions (modes) $\{\hat{\theta}_n\}_{n=1}^N$ due to independent initializations and training dynamics, while MC Dropout defines a local variational distribution around each configuration by keeping dropout active at test time. This motivates the approximation
\begin{equation}
    q(\theta) = \frac{1}{N}\sum_{n=1}^{N} q_n(\theta).
\end{equation}
with corresponding predictive approximation

\begin{align}
p(y^* \mid x^*, \mathcal{D})
&\approx \int p(y^* \mid x^*, \theta)\, q(\theta)\, d\theta \\
&= \frac{1}{N}\sum_{n=1}^{N}\int p(y^* \mid x^*, \theta)\, q_n(\theta)\, d\theta \\
&\approx \frac{1}{N}\sum_{n=1}^{N}\mathbb{E}_{z \sim q(z)}
\big[p(y^* \mid x^*, \hat{\theta}_n, z)\big] \\
&\approx \frac{1}{NM}\sum_{n=1}^{N}\sum_{m=1}^{M}
p\big(y^* \mid x^*, \hat{\theta}_n, z_{n,m}\big),
\end{align}
where $z_{n,m} \sim q(z)$ denotes a dropout mask (sampled independently for each
network $n$ and Monte Carlo pass $m$).

%
%
%
%
%
\begin{figure*}[h]
    \centering
    \includegraphics[width=0.98\textwidth]{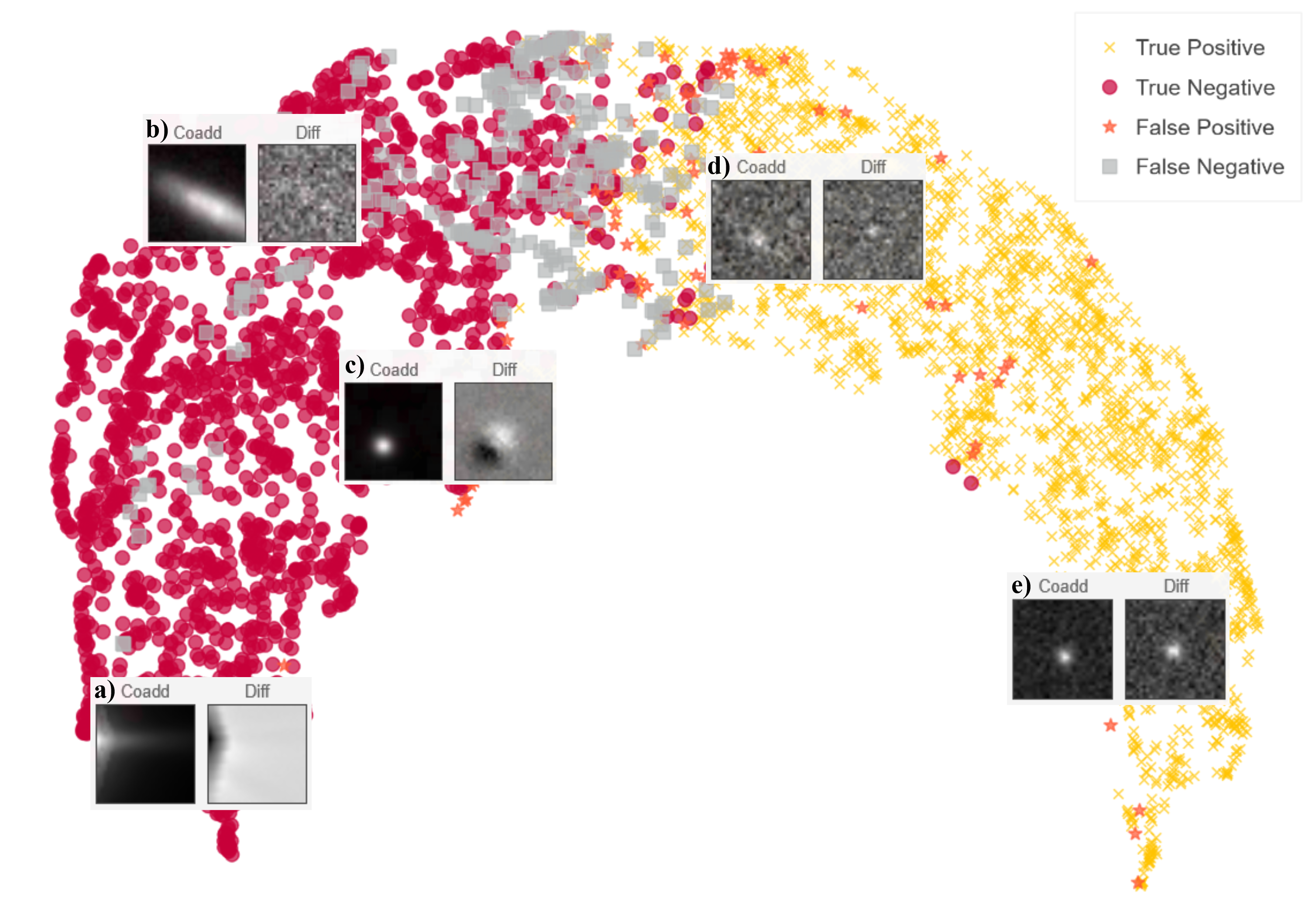}
    \caption{
    Two-dimensional UMAP projection of the embeddings from the penultimate dense layer of the classifier. Each point corresponds to one candidate and is colored by outcome relative to the labels: TP (yellow $\times$), TN (red $\bullet$), FP (orange $\star$), and FN (gray $\blacksquare$). Insets (a-e) show representative Coadd (template) and Diff (difference) cutouts sampled from the indicated regions: (a) negative-Diff region, dominated by negative residuals characteristic of subtraction artifacts or over-subtraction; (b) low-SNR region, where candidates are difficult to distinguish from background fluctuations; (c) dipole region, showing positive/negative residual astrometric misalignment; (d) low-SNR transient region, containing faint point-source-like residuals consistent with marginal transient detections; and (e) high-SNR transient region, containing compact, symmetric positive residuals characteristic of confident transient-like candidates. The cutouts are provided as illustrative examples only, they were assembled manually to make the static figure informative and are therefore approximate with respect to exact UMAP locations and local neighborhoods. The interactive versions of the UMAP visualizations (with different overlays) are available in the Zenodo supplementary material (DOI:\href{https://doi.org/10.5281/zenodo.18434676}{10.5281/zenodo.18434676}).
    }
    \label{fig:umap_latent}

\end{figure*}
\section{Latent Space Representation}
\label{sec:latent}
Analysis of learned internal representations can provide useful insight into model behavior. In this work, we study the latent representation learned by the classifier by extracting hidden activations and visualizing them in two dimensions. This analysis is intended as a global representation-level probe of the model, rather than a full mechanistic interpretation of its internal computations \citep{sharkey2025openproblemsmechanisticinterpretability}.

While much recent work focuses on identifying interpretable concepts in latent representations using sparse autoencoders (SAEs) \citep{SAE_galaxy_2025}, we adopt a simpler approach based on direct exploration of latent activations through nonlinear dimensionality reduction.

\subsection{Feature Extraction}

The dimensionality of the latent representation depends on the layer under consideration. We focus on the output of the final hidden layer of the network (Fig.~\ref{fig:CNN-architecture}), where we expect high-level abstractions and task-relevant patterns to be strongly represented.

For each input sample, we record the activation vector at this layer. For a dataset of $N$ samples, this gives a feature matrix $\mathbf{F} \in \mathbb{R}^{N \times d},$ where $d$ denotes the dimensionality of the selected layer's output.

When $N$ exceeds a predefined maximum sample size $n_{\text{max}}$, we apply stratified random sampling to obtain a representative subset while preserving the class distribution.
For ensemble-based and co-teaching models, features are extracted from the first model in the ensemble. 

\subsection{Dimensionality Reduction with UMAP}
The high dimensionality of $\mathbf{F}$ makes direct interpretation infeasible. We therefore use Uniform Manifold Approximation and Projection (UMAP) \citep{UMAP2020} to compute a two-dimensional embedding suitable for visualization.

UMAP constructs a weighted nearest-neighbor graph in the original feature space and optimizes a low-dimensional embedding that preserves local neighborhood structure. As with other nonlinear embedding methods, the visualization is primarily useful for exploring local organization; global distances in the 2D projection should not be over-interpreted.

In contrast to linear projections such as PCA \citep{PCA_1987}, UMAP can capture nonlinear structure; similarly to t-SNE \citep{tsne2008}, it primarily emphasizes preservation of local neighborhood relationships. In practice, UMAP often produces embeddings with more apparent global continuity than t-SNE, although global distances are not guaranteed to be preserved.

We compute a two-dimensional UMAP embedding with \texttt{n\_components}=2 and Euclidean distance in the original feature space. We vary \texttt{n\_neighbors} and \texttt{min\_dist} over a small grid:
\begin{itemize}
    \item \textbf{n\_neighbors}: size of the local neighborhood used to approximate the manifold,
    \item \textbf{min\_dist}: minimum distance between embedded points, controlling cluster compactness,
    \item \textbf{n\_components}: target embedding dimensionality (set to 2),
    \item \textbf{metric}: Euclidean distance in the original feature space.
\end{itemize}
Specifically, we perform a grid search over
\begin{align*}
    \text{n\_neighbors} \in \{5, 10, 15, 30, 50\}, \\
    \text{min\_dist} \in \{0.0, 0.1, 0.25, 0.5\}.
\end{align*}
and select the configuration using a simple visualization heuristic based on the spread of the embedding coordinates. This heuristic is used only to avoid overly collapsed visualizations and should not be interpreted as an objective measure of representation quality.

Using an interactive graph built with the Bokeh library, we visualize the UMAP representation together with the corresponding confusion-matrix categories, as well as image SNR, predicted class probability, and model uncertainty when available. This interactive tool facilitates qualitative exploration of model behavior and helps identify structured failure modes or mislabeled examples.
The resulting embeddings are used as a qualitative analysis tool to inspect class structure, misclassifications, uncertainty patterns, and potential label noise in the learned representation.

%
%
%
%
%
\section{Results}
\label{sec:results}

\subsection{Weakly supervised learning}
Table~\ref{tab:wsl_results} compares weakly supervised learning (WSL) approaches under increasing label-noise rates on training sets, reporting subtype-specific recall for the positive class (survey transients) and specificity for the negative class (bogus detections), together with aggregate accuracy and receiver operating characteristic (ROC). 

When trained on the baseline dataset (current HSC injected survey data), all methods achieve strong performance (Acc $\approx 0.91$-$0.92$; ROC $\approx 0.97$), with Co-teaching and Asym-Co-teaching providing the strongest overall balance across metrics. 

As we artificially increase the label noise in the baseline, the Standard model degrades substantially (e.g., at 35\% noise, Acc $=0.812$ and ROC $=0.903$), whereas the co-teaching variants remain markedly more robust. In particular, Co-teaching and Asym-Co-teaching consistently maintain high ROC AUC across noise levels ($\approx 0.95$) and improve accuracy relative to the Standard approach.  

At higher noise (25-35\%), Asym-Co-teaching offers the best aggregate results (Acc $\approx 0.87$-$0.89$; ROC $\approx 0.95$), indicating that co-teaching, especially with asymmetric handling of label noise, provides the most stable performance as annotation quality deteriorates.

The comparable recall obtained for the SN-like and OT subclasses supports the hypothesis that, at the single-image level, SN-like injections provide useful training examples for detecting other transient classes as well.
In addition to the two methods presented in Section~\ref{sec:wsl}, we also evaluated Stochastic Co-teaching \citep{stocoteaching_vos2023}, which we briefly describe in Appendix~\ref{app:coteaching-form}. It exhibits pronounced trade-offs across the noise regimes. This instability was observed throughout training, with difficulty in tuning the $\alpha$ and $\beta$ parameters and in selecting the delay before rejection begins. In most training cases, Stochastic Co-teaching collapsed to rejecting predominantly one class, leading to unstable training.

\begin{table}[h]
\centering
\small
\setlength{\tabcolsep}{6pt}
\renewcommand{\arraystretch}{1.2}
\begin{tabular}{ll|ccccc}
\hline
\textbf{Noise} & \textbf{WSL method} &
\textbf{SN-like} & \textbf{OT} & \textbf{B} & \textbf{Acc} & \textbf{ROC} \\
\hline
\multirow{3}{*}{B}
 & Standard                 & 0.885 & 0.914 & 0.926 & 0.912 & 0.966 \\
 & Co-t                     & 0.919 & 0.928 & 0.915 & 0.920 & 0.972 \\
 & \textbf{Asym-Co-t }      & 0.898 & 0.905 & 0.944 & 0.922 & 0.971 \\

\hline
\multirow{3}{*}{15\%}
 & Standard               & 0.798 & 0.815 & 0.934 & 0.866 & 0.946 \\
 & Co-t                   & 0.920 & 0.925 & 0.872 & 0.899 & 0.963 \\
 & \textbf{Asym-Co-t }    & 0.843 & 0.838 & 0.932 & 0.883 & 0.959 \\

\hline
\multirow{3}{*}{25\%}
 & Standard               & 0.792 & 0.753 & 0.937 & 0.849 & 0.939 \\
 & Co-t                   & 0.882 & 0.897 & 0.899 & 0.894 & 0.959 \\
 & \textbf{Asym-Co-t }    & 0.887 & 0.875 & 0.903 & 0.891 & 0.951 \\

\hline
\multirow{3}{*}{35\%}
 & Standard               & 0.764 & 0.761 & 0.868 & 0.812 & 0.903 \\
 & Co-t                   & 0.807 & 0.779 & 0.876 & 0.831 & 0.921 \\
 & \textbf{Asym-Co-t }    & 0.812 & 0.833 & 0.925 & 0.871 & 0.950 \\

\hline
\end{tabular}
\caption{Evaluation of weak supervision learning methods under varying label noise conditions. For the positive class (survey transients), recall is reported separately for Supernova like (SN-like) and other transient or variable types (OT). For the negative class, specificity is shown for bogus detections (B). Overall accuracy (Acc) and ROC AUC summarize aggregate performance.}
\label{tab:wsl_results}
\end{table}

\subsection{Uncertainty Quantification}
\label{sec:results-uq}

\begin{figure}[h]
    \centering
    \includegraphics[width=1\linewidth]{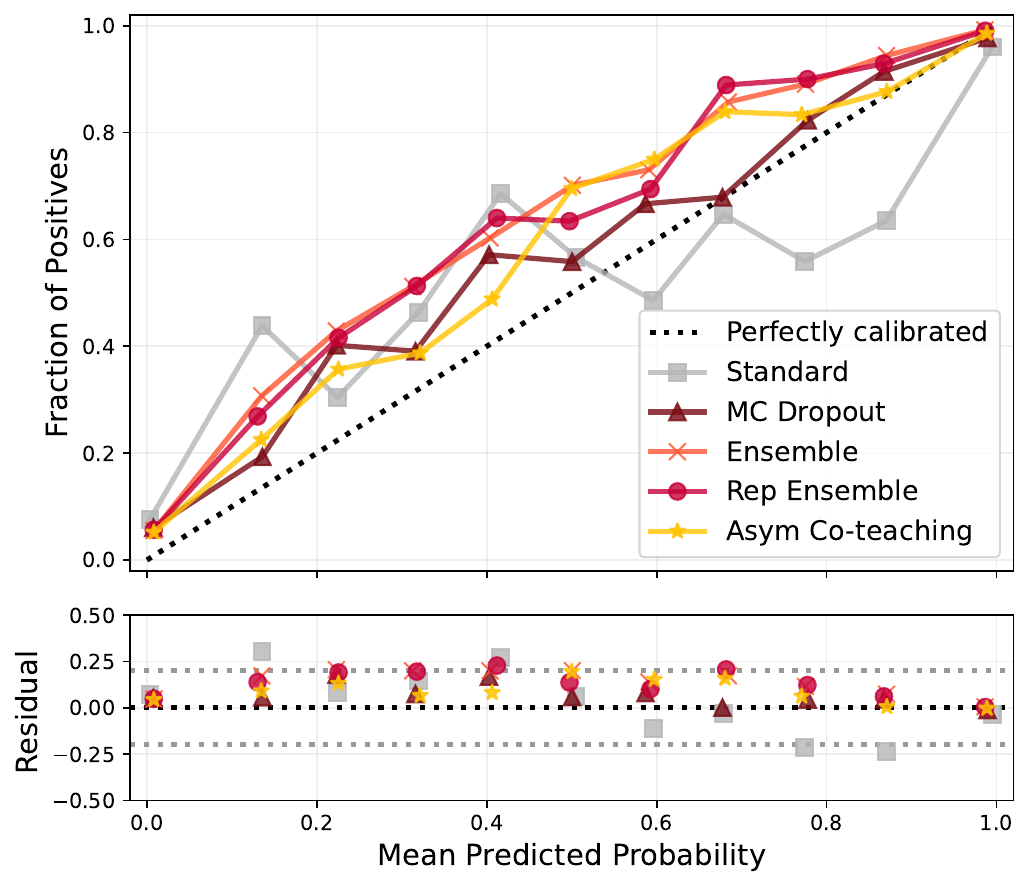}
    \caption{The top panel shows the fraction of positive samples as a function of the mean predicted probability, using $11$ equally spaced probability bins. The dashed diagonal indicates perfect calibration. The bottom panel shows the residuals with respect to perfect calibration.}
    \label{fig:distrib_calibration}
\end{figure}

Table~\ref{tab:uq_calibration} summarizes probabilistic performance and calibration for the different uncertainty-quantification strategies. All UQ approaches improve over the standard deterministic baseline, with consistent reductions in NLL, Brier score, and ECE. The standard CNN achieves an NLL of $0.453$ and a Brier score of $0.074$, illustrating the comparatively poorer probabilistic quality of deterministic predictions. This behavior is easily noticeable in Fig.~\ref{fig:distrib_calibration}.

Our method, Ensemble-MC-Dropout for Co-teaching (Section~\ref{sec:unc_asy_coteaching}) achieves the best overall performance across the three reported metrics, with the lowest NLL ($0.292$), the lowest Brier score ($0.063$), and the lowest ECE ($0.0375$). Deep Ensembles and Repulsive Ensembles remain competitive, particularly on NLL and Brier score, but are consistently outperformed by our approach, with the clearest advantage observed in ECE.
Relative to the standard model, this corresponds to a $35.5\%$ reduction in NLL, a $14.9\%$ reduction in Brier score, and a $40.8\%$ reduction in ECE, indicating both improved probabilistic accuracy and substantially better calibration. 
These results indicate that our Ensemble-MC-Dropout inference strategy, combined with dual-network architecture, produces probability estimates that more accurately reflect true class frequencies. The lower NLL, in particular, demonstrates that our method effectively reduces overconfident incorrect predictions, a critical property for downstream tasks. The simultaneous improvement in three metrics confirms that our uncertainty quantification approach achieves superior calibration without sacrificing robustness to extreme probabilities.

\begin{table}[h]
\centering
\small
\setlength{\tabcolsep}{6pt}
\renewcommand{\arraystretch}{1.2}
\begin{tabular}{l|ccc}
\hline
\textbf{Metric} & \textbf{NLL} & \textbf{Brier Score} & \textbf{ECE} \\
\hline
Standard             & $0.453$ & $0.074$ & $0.0633$\\
MC Dropout           & $0.334$ & $0.069$ & $0.0405$\\
Deep Ensemble        & $0.314$ & $0.065$ & $0.0535$\\
Repulsive Ensemble   & $0.308$ & $0.065$ & $0.0511$\\
\textbf{Our method}  & $0.292$ & $0.063$ & $0.0375$\\
\hline
\end{tabular}
\caption{Negative Log-Likelihood (NLL), Brier Score and Expected Calibration Error (ECE) on $11$ bins for the standard model, the Deep Ensemble, Repulsive Ensemble and our method, Ensemble-MC-Dropout for Co-teaching}
\label{tab:uq_calibration}
\end{table}

In addition to calibration, a useful uncertainty estimator should assign larger uncertainty to samples that are intrinsically more difficult to classify, such as low-SNR detections or supernovae observed far from peak brightness. Table~\ref{tab:uq_results} reports Spearman correlations between uncertainty and these two physically meaningful quantities.

All methods show the expected negative correlation with SNR ($\rho < 0$), indicating that uncertainty increases as signal quality degrades. Our method achieves the strongest correlation ($\rho = -0.296$), marginally outperforming Deep Ensembles ($\rho = -0.290$) and substantially improving upon MC Dropout ($\rho = -0.239$). 

All methods also exhibit the expected positive correlation with distance from maximum brightness. Here, Deep Ensembles achieve the highest correlation ($\rho=0.221$), with our method performing nearly identically ($\rho=0.218$).

Overall, these results show that the proposed method produces uncertainty quantification that are physically meaningful and competitive with substantially more expensive ensemble-based baselines, while requiring only two jointly trained networks augmented with MC Dropout at inference time, which greatly reduces training cost compared with large ensembles.

\begin{table}[h]
\centering
\small
\setlength{\tabcolsep}{6pt}
\renewcommand{\arraystretch}{1.2}
\begin{tabular}{l|cc}
\hline
\textbf{Physical value} & $\boldsymbol{\rho}$ \textbf{SNR} & $\boldsymbol{\rho}$ \textbf{Max. bright} \\
\hline
MC Dropout           & $-0.239$ & $0.176$ \\
Deep Ensemble        & $-0.290$ & $0.221$ \\
Repulsive Ensemble   & $-0.271$ & $0.198$ \\
\textbf{Our method}  & $-0.296$ & $0.218$ \\
\hline
\end{tabular}
\caption{Spearman rank correlation coefficients ($\rho$) between predicted uncertainty and signal-to-noise ratio (SNR) and distance from maximum brightness (Max. bright, in days). Negative $\rho_{\text{SNR}}$ values indicate that uncertainty appropriately increases for low-SNR detections, while positive $\rho_{\text{Max. bright}}$ values indicate higher uncertainty for supernovae observed far from peak luminosity.}
\label{tab:uq_results}
\end{table}
It is also informative to examine the class-conditional values of $\rho_{\text{SNR}}$, reported in Appendix~\ref{app:uq-snr-perclass}. The overall correlation between uncertainty and SNR is driven primarily by the true-positive subset. A weaker but still visible trend remains in the true-negative subset, whereas the correlation is much less pronounced for misclassified examples.

This suggests that SNR is most directly informative for uncertainty within the transient-like detections, where low signal quality more strongly affects classification confidence. This interpretation is consistent with the class-dependent SNR distributions shown in Appendix~\ref{app:SNR_app}, and with the latent-space visualization in App.~\ref{app:UMAP-overlay} (interactive version available on Zenodo), where low-SNR samples appear more diffusely distributed within the transient region than in the bogus region.


\subsection{Latent space representation }

The projection displayed in Fig.~\ref{fig:umap_latent} highlights a broad separation between negative and positive manifolds, with most ambiguous cases and misclassifications concentrated near their interface, thereby shaping the decision boundary.

One key finding of the UMAP representation of the latent space is that it allows the user to understand which patterns are being identified by the model even within a class. In addition to the class information, UQ and SNR can be displayed (see Appendix~\ref{app:UMAP-overlay}). 
  
The visualization is also useful for inspecting misclassified cutouts and identifying structured failure modes or potential label noise. We therefore use it as a qualitative complement to the quantitative analyses, rather than as a standalone measure of model performance.

\subsection{Towards object identification from source classifications}
\label{sec:results-lc}

\begin{figure*}[t]
    \centering
    \includegraphics[width=1\linewidth]{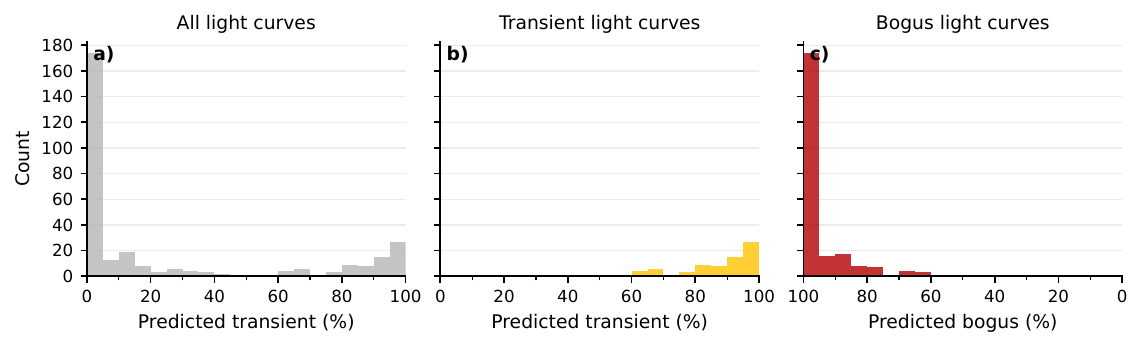}
    \caption{Light-curve-level classification score distributions. Shown are the distributions of the fraction of source detections classified as transient for all light curves \textbf{(a)} and for true transient light curves \textbf{(b)}, together with the distribution of the fraction of source detections classified as bogus for true bogus light curves \textbf{(c)}.}
    \label{fig:lc_source_distrib}
\end{figure*}

Although the model classifies DIA sources independently, its predictions can be grouped at the object level to perform light-curve-level transient-bogus discrimination. Using the method described in Section~\ref{sec:unc_asy_coteaching}, we aggregate source-level predictions within each labeled light curve and assign an object-level label by majority vote. The resulting performance is summarized in Table~\ref{tab:lc_id_phase_summary}.

Figure~\ref{fig:lc_source_distrib} shows the distribution, for each light curve in the labeled evaluation set, of the fraction of sources classified as transient; the corresponding cumulative counts are shown in App.~\ref{app:cumul-lightcurve-count}. A first notable result is the strong internal consistency of the source-level predictions within individual light curves: $232/306$ light curves ($75.8\%$) have at least $90\%$ of their sources assigned to the correct class. For the transient class specifically, $79.7\%$ of light curves have more than $80\%$ of their sources classified as transient.

On this curated evaluation set, aggregation of the source-level predictions yields very strong object-level performance: $304/306$ light curves are assigned the correct binary label ($99.35\%$ accuracy), with only two errors overall (one false positive and one false negative). At the source level, class separation also remains strong, with similar recall for supernovae and other transients or variables ($0.887$ and $0.896$, respectively) and a bogus specificity of $0.951$.

The uncertainty quantification remains consistent with observational difficulty. For supernova detections, uncertainty is lowest near maximum light (mean uncertainty $\simeq 0.074$) and increases substantially for observations far from peak ($>50$ days; mean uncertainty $\simeq 0.180$), where the corresponding accuracy falls to $0.75$. This behavior is consistent with the uncertainty analysis presented above and suggests that epoch-level uncertainty could be useful for down-weighting ambiguous, low-information measurements in downstream light-curve analyses.

Together, these results indicate that single-epoch, per-detection classification scores (and their associated uncertainties) are sufficiently stable to support reliable light-curve identification after grouping, while also providing a principled signal for down-weighting ambiguous, low-information epochs during downstream light-curve-based analyses.

Encouraged by the strong performance obtained on the labeled set (Table~\ref{tab:lc_id_phase_summary}), we then deploy the same grouping rule as an inference-time filter on the full UDEEP dataset, which contains $3,856,539$ distinct objects. Among these, $269,968$ objects have more than $50\%$ of their sources classified as transient. However, a more detailed inspection shows that most of these objects are observed on only a single night. Requiring at least five different nights of observation reduces this set to $1,770$ transient candidates, including $108$ with more than $50$ distinct observations.

%
%
%
%
%
\section{Discussion and open questions}
\label{sec:discussion}

The aim of this work was to address the two challenges introduced in Section~\ref{sec:intro}: learning real-bogus classification without human-labeled training data, and obtaining uncertainty quantification and representations that remain interpretable and useful for downstream analysis.
Our experiments show that the proposed framework recovers strong source-level performance and, after grouping detections, very strong object-level performance on the manually curated evaluation set (Section~\ref{sec:results-lc}).

Although the single-source metric may appear less reliable, it is important to note that this limitation is not solely algorithmic. Indeed, our labels are derived from observed light curves, and a fraction of events is intrinsically ambiguous given the available observables. This sets an effective upper bound on achievable performance for the single-source task. Inspection of misclassified cutouts with the interactive visualization supports this interpretation. Combined with in-depth latent-space visualization, these results help explain the observed behavior.    

A key strength of the method is its portability across surveys with limited adaptation. In the context of forthcoming datasets such as LSST, it remains an open question whether direct inference on survey data, despite the expected domain shift, is sufficient, or whether performance and calibration are best preserved by re-running the full pipeline with injection, training, and inference using survey-specific simulations. In both cases, the methodology is survey-agnostic, with the main challenges being adaptation to other data-management tools, as we have shown that our method is robust to extreme noise rates.

Beyond the present scope, weakly supervised methods appear promising both for our current approach, fully human label-free training, and as a way to mitigate the inherent label noise expected in community-labeled datasets. 
In parallel, asymmetric co-teaching could potentially be leveraged to estimate dataset noise levels, for example by exploiting the training dynamics to infer class-dependent noise rates and, in turn, a robust Real-Bogus ratio. While noise-rate estimation is discussed in \citet{stocoteaching_vos2023} for a global noise level, extending it to class-conditional noise in survey data is an interesting direction for future work.

Regarding uncertainty quantification (UQ), it is noteworthy that our approach matches or outperforms standard baselines, including MC dropout and large ensembles. Given that ensembles with $N=50$ are expected to explore a broader area of the loss landscape than methods that differ mainly at inference time \citep{fort2020deepensembleslosslandscape}, this result suggests that the training coupling induced by Asym-Co-teaching may play a central role. An important question for future work is whether the gain arises primarily from Asym-Co-teaching and the interaction between the two models during training, or whether the combination of MC dropout and a small ensemble could serve as a robust alternative to large ensembles. 

Finally, we find that UQ correlates with physically motivated quantities and improves calibration.
Consistent with this, latent-space projections indicate that the UQ captures the expected decision boundary, whereas SNR is less structured in representation space, particularly in the bogus region. This supports the hypothesis that the model relies on features that are not well summarized by SNR alone, motivating further work to identify the dominant image and contextual factors driving both predictions and uncertainties.

In addition, it would be valuable to explore concept-based interpretability methods applied to the learned representations \citep{tcav_kim_2018}, as well as more advanced methods such as \citep{SAE_galaxy_2025}, to identify human-interpretable concepts associated with both astrophysical transients and bogus detections. Such concepts may guide targeted diagnostics, improve the realism of injections, and potentially lead to explicit bogus simulation. 

\section{Conclusion}
\label{sec:conclusion}
We presented a human-label-free approach to Real-Bogus classification for transient candidates. The method is trained using physically motivated injected transient classes together with a highly contaminated survey class, enabling learning without relying on human-labeled training data. This setup allows us to build on weakly supervised learning and to develop a training strategy dedicated to an asymmetric-noise regime. We show that the approach remains robust under noise levels exceeding the transient-to-bogus ratio typical of current pipelines, supporting its applicability to future surveys.

We also presented a comparative study of uncertainty quantification methods using both calibration and physically motivated metrics. Building on the dual-network setting of co-teaching, we introduced a hybrid strategy combining elements of deep ensembles and MC dropout. This approach reduces computational cost while achieving UQ performance comparable to, or better than, the state-of-the-art baselines considered here. 
To support interpretation, we developed a visualization framework enabling in-depth analysis of model behavior. Finally, we extended our evaluation to light-curve space, where we obtain near-perfect classification accuracy on the labeled set.

\vspace{0.5cm}
{\small
\noindent\textit{Author contributions.} 
R.B.G. contributed to the data processing with the LSST Gen3 pipeline, and carried out the methodological development, including the physically motivated injection strategy, weakly supervised learning framework, and uncertainty quantification methods. He performed the main analysis, implemented the \textit{ML4transients} library, and wrote the manuscript.
B.S. provided expertise on the LSST Gen2 and Gen3 difference-imaging pipeline, including resolving compatibility issues, supporting database management, and supervising the project.
D.F. conceived the project, initiated the data-processing strategy and latent-space analysis, and provided overall scientific supervision.
B.R. initiated the analysis of light curves and difference-image cutouts and implemented an initial CNN model for real–bogus classification.
M.Y. led the initial data-processing effort, running the LSST Gen2 pipeline for HSC and developing an early source-injection framework.
M.S. further developed and systematically evaluated the CNN, including performance studies and the incorporation of additional inputs.
M.G. improved the CNN through hyperparameter optimization and developed the UMAP-based analysis and visualization tools.
V.P. provided guidance on interpretability.}
\begin{acknowledgements}
      Part of this work was supported by the European Union’s Horizon Europe research and innovation programme under the Marie Sklodowska-Curie grant agreement No 101168829, Challenging AI with Challenges from Physics: How to solve fundamental problems in Physics by AI and vice versa (AIPHY). 
\end{acknowledgements}

\bibliographystyle{aa}
\bibliography{astro_ref,ML_ref}

\begin{appendix}

\section{Injection magnitude}
\label{app:magnitude}

\noindent
We compare the magnitude distributions of injected sources, recovered injections, and real detections with positive flux. As expected, the recovered injections are systematically brighter than the full injected population, reflecting the detection efficiency of the pipeline at faint magnitudes. The positive-flux real detections have a broadly comparable magnitude distribution, and the resulting sample remains approximately balanced between real and recovered injected detections.

\begin{figure}
  \centering
  \includegraphics[width=0.8\columnwidth]{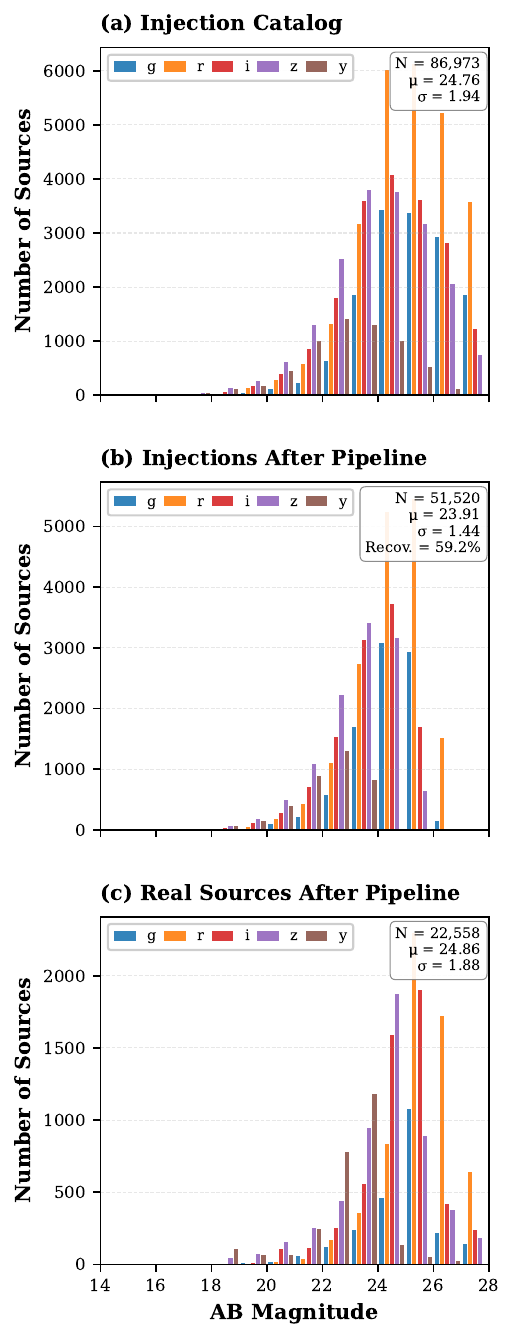}
  \caption{Magnitude distributions for (a) injected sources, (b) recovered injections after the pipeline, and (c) real sources with positive flux. In each panel we report the number of sources $N$ and the best-fitting Gaussian parameters ($\mu$, $\sigma$) describing the distribution.}
  \label{fig:mag_distrib}
\end{figure}

\clearpage

\section{Light curve labeling tool}
\label{app:lc-tool}
Examples of the actual labeling tool are available in the Zenodo supplementary material (DOI: \href{https://doi.org/10.5281/zenodo.18434676}{10.5281/zenodo.18434676}).
\begin{figure}[ht]
    \centering
    \includegraphics[width=1\linewidth]{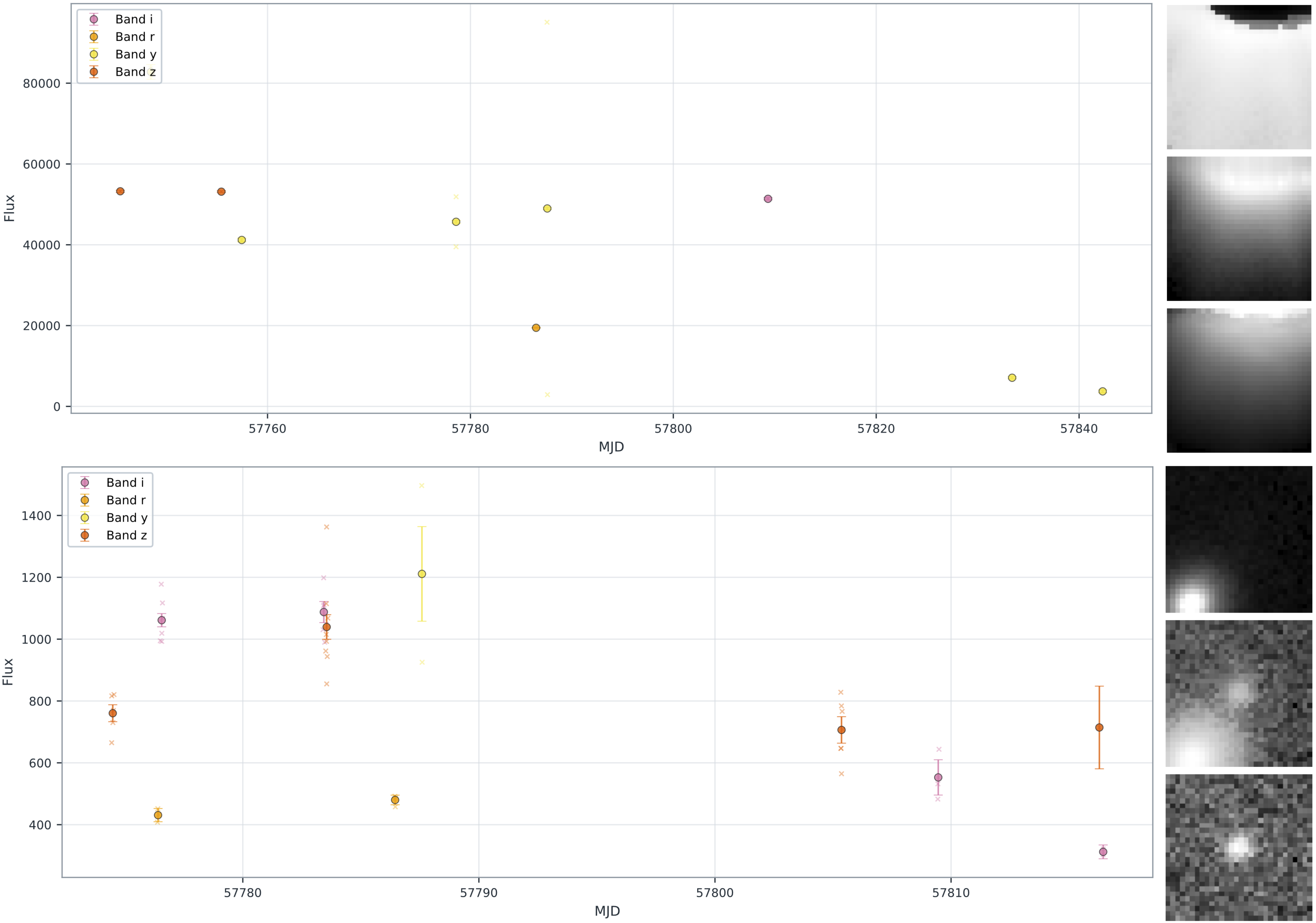}
    \caption{Representative examples of a bogus candidate (top) and a supernova-like transient (bottom). Left: multi-band light curves in detected bands as a function of MJD. Right: examples of associated cutouts shown. For each object from top to bottom: coadd image, mean of the science for this band and mean of the difference image for the band.}
  \label{fig:lc}
\end{figure}
\section{Training configurations}
\label{app:param}
For reproducibility, we summarize below the training hyperparameters used for each uncertainty-estimation or robust-training method considered in this work. Unless stated otherwise, all settings follow the default values described in the main text, and we report only the parameters specific to each method (number of models or forward passes, repulsion strength and type, dropout probabilities, and the co-teaching forget-rate schedules). For the forget rates, the reported values correspond to the $\{\text{Baseline}, 15\%, 25\%, 35\%\}$ sets.

\begin{itemize}\setlength\itemsep{0.2em}
  \item \textbf{Ensemble:} $\texttt{num\_models}=50$.
  \item \textbf{Repulsive ensemble:} $\texttt{num\_models}=50$, $\lambda_{\mathrm{rep}}=0.1$,
        $\texttt{repulsion\_type}=\texttt{"prediction"}$, $T=1.0$.
  \item \textbf{MC Dropout:} $\texttt{fwd\_passes}=50$,
        $p_{\mathrm{dense}}=0.25$, $p_{\mathrm{conv}}=0.125$.
  \item \textbf{Co-teaching:} $\texttt{forget\_rate}\in\{0.06,0.16,0.26,0.36\}$.
  \item \textbf{Asym. Co-teaching:} $(\texttt{forget\_rate}_0,\texttt{forget\_rate}_1)\in
        \{(0.05,0.01),(0.15,0.01),(0.25,0.01),(0.35,0.01)\}$.
\end{itemize}
We select the forget-rate schedules using prior knowledge estimated from the label noise inferred by the standard training method. In particular, we use the observed false-positive and false-negative fractions on the training set as approximate indicators of mislabeling. This suggests a noise level of about $5\%$ for the survey (baseline) class and about $1\%$ for the injected class; the reported forget rates are chosen to reflect these estimates.

\section{WSL formalism}
\label{app:coteaching-form}

\subsection{Co-teaching}

Formally, let $\mathcal{B}_t = \{(x_i, \tilde  y_i)\}_{i \in I_t}$ denote the mini-batch at iteration (or epoch) $t$, where $I_t$ is the set of indices of the samples in the batch and $|\mathcal{B}_t| = |I_t|$.

The per-sample losses of network~1 and network~2 on this batch are defined by their Binary Cross Entropy ($\mathbf{BCE}$) as
\begin{equation*}
\ell_i^{(1)}(t) = \mathbf{BCE}\big(f_{\theta_1}(x_i), \tilde y_i\big), 
\qquad
\ell_i^{(2)}(t) = \mathbf{BCE}\big(f_{\theta_2}(x_i), \tilde y_i\big),
\quad i \in I_t.
\end{equation*}

The forget rate is defined as
\begin{equation}
r(t) = \min\left(r \cdot \frac{t}{T_k},\, r\right),
\end{equation}
where $T_k$ is a predefined epoch after which the forget rate remains constant. Given the forget rate, the fraction of samples kept in the batch at iteration $t$ is $1 - r(t)$. We set
\begin{equation*}
k_t = \big\lfloor (1 - r(t))\,|\mathcal{B}_t| \big\rfloor
    = \big\lfloor (1 - r(t))\,|I_t| \big\rfloor.
\end{equation*}

For network~1, we sort the indices in $I_t$ according to the increasing order of its loss:
\begin{equation*}
\ell_{i_1^t}^{(1)}(t) \le \ell_{i_2^t}^{(1)}(t) \le \cdots \le \ell_{i_{|I_t|}^t}^{(1)}(t),
\end{equation*}
where $(i_1^t,\dots,i_{|I_t|}^t)$ is a permutation of $I_t$.

The set of reliable samples selected by network~1 at iteration $t$ is then defined as
\begin{equation*}
\mathcal{R}_{1,t} 
= \{ i_1^t, \dots, i_{k_t}^t \} 
= \arg\min_{\substack{S \subset I_t \\ |S| = k_t}}
    \sum_{i \in S} \ell_i^{(1)}(t).
\end{equation*}

The Co-teaching losses are then given by
\begin{equation}
\begin{aligned}
\mathcal{L}_{\text{co-teach}}^{(1)}(t) 
&= \frac{1}{|\mathcal{R}_{2,t}|}
  \sum_{i \in \mathcal{R}_{2,t}} 
  \mathbf{BCE}\big(f_{\theta_1}(x_i),\tilde y_i\big),
\\[0.3em]
\mathcal{L}_{\text{co-teach}}^{(2)}(t) 
&= \frac{1}{|\mathcal{R}_{1,t}|}
  \sum_{i \in \mathcal{R}_{1,t}} 
  \mathbf{BCE}\big(f_{\theta_2}(x_i),\tilde y_i\big),
\end{aligned}
\end{equation}
where $\mathcal{R}_{1,t}$ and $\mathcal{R}_{2,t}$ denote the sets of reliable samples selected at iteration $t$ by networks~1 and~2, respectively.
\subsection{Stochastic Co-teaching}
Standard Co-teaching requires the user to specify a forget rate in advance. Stochastic Co-teaching \citep{stocoteaching_vos2023} replaces this fixed rate with a stochastic rejection rule. At iteration \(t\), each network samples a threshold
\begin{equation*}
    z_t \sim \mathrm{Beta}(\alpha,\beta),
\end{equation*}
where \(\alpha\) and \(\beta\) control the expected strictness of sample rejection.

For a mini-batch \(\mathcal{B}_t=\{(x_i,\tilde y_i)\}_{i\in I_t}\), network \(j\in\{1,2\}\) outputs a probability
\begin{equation*}
    \hat p_i^{(j)}(t)=\sigma\!\bigl(f_{\theta_j}(x_i)\bigr).
\end{equation*}

The probability assigned to the observed training label is

\begin{align*}
p_i^{(j)}(t)=
\begin{cases}
\hat p_i^{(j)}(t), & \text{if } \tilde y_i=1,\\
1-\hat p_i^{(j)}(t), & \text{if } \tilde y_i=0.
\end{cases}
\end{align*}

Samples with \(p_i^{(j)}(t)\le z_t\) are rejected, and the remaining samples are passed to the peer network for updating, as in standard Co-teaching.
This stochastic rule removes the need to fix a single deterministic forget rate, but in our experiments it remained difficult to tune and often led to unstable class-wise rejection behavior. For this reason, we report it only as an auxiliary baseline and do not retain it among the main methods.

\section{Signal-to-noise ratio by class}
\subsection{Signal-to-noise ratio distribution}
\label{app:SNR_app}
\begin{figure}[h]
    \centering
    \includegraphics[width=0.9\columnwidth]{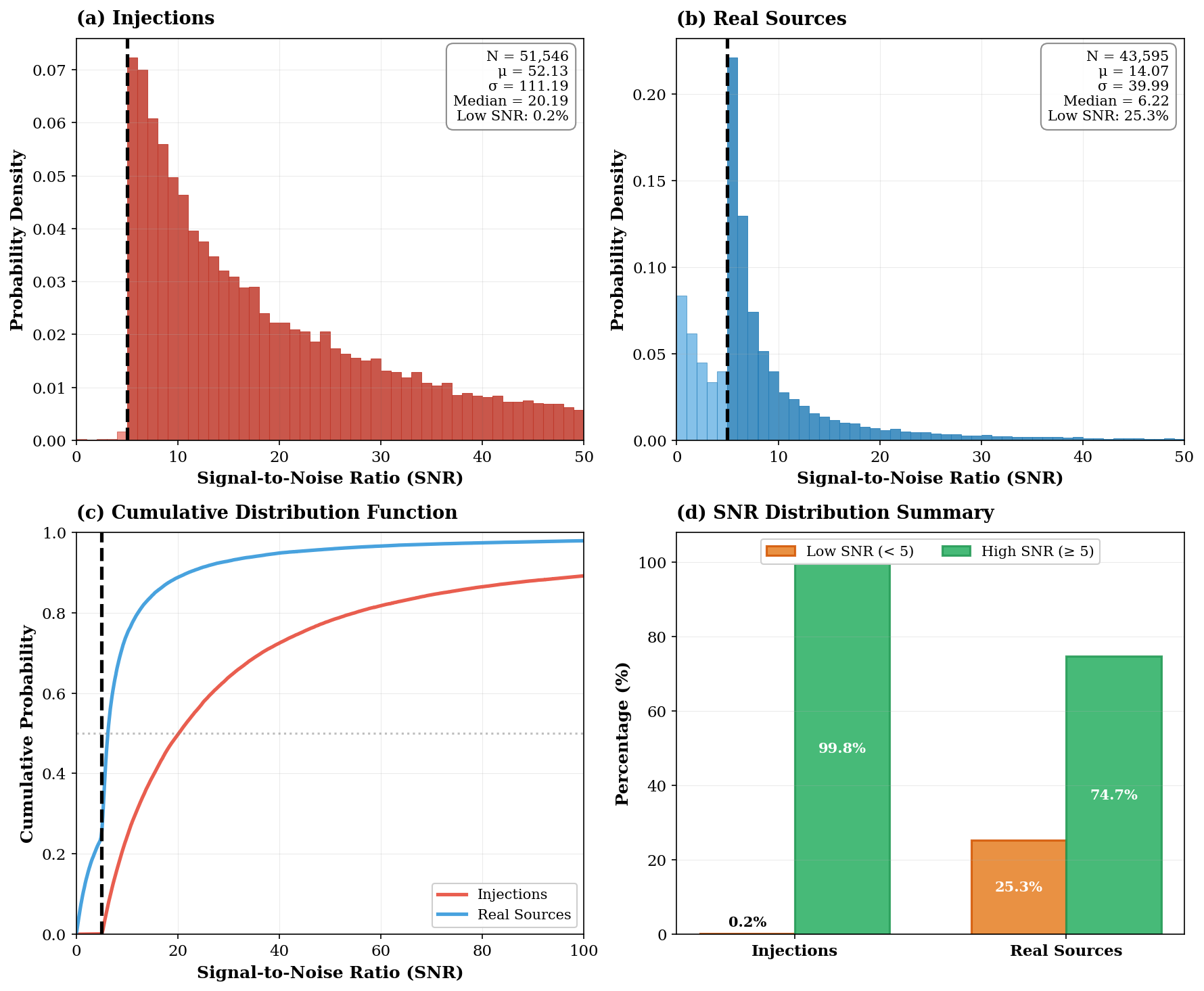}
    \caption{Signal-to-noise ratio analysis of the \textit{rc2\_subset} dataset by class. Panel (a) shows the SNR distribution of the injected class, panel (b) the SNR distribution of the survey class, panel (c) the corresponding cumulative distribution functions, and panel (d) the distribution split below and above $\mathrm{SNR}=5$.}
    \label{fig:SNR_class}
\end{figure}

The SNR is computed as: 
\begin{equation*}
    \mathrm{SNR}=\frac{|\textit{psfFlux}|}{\textit{psfFluxErr}},
\end{equation*}

where \textit{psfFlux} is the PSF-fitted flux measured on the difference image and \textit{psfFluxErr} is its associated uncertainty.
Figure~\ref{fig:SNR_class} shows that the SNR distributions differ substantially between the two training classes. In particular, detections with $\mathrm{SNR}\leq5$ are nearly absent from the injected class because such faint events are less likely to be recovered by the detection pipeline. The injected class also extends to systematically higher SNR values than the survey class. This class asymmetry is expected from the construction of the dataset, but it should be kept in mind when interpreting the relationship between SNR and uncertainty.

\subsection{Signal-Noise-Ratio correlation with Uncertainty Quantification per class}
\label{app:uq-snr-perclass}
Table~\ref{tab:uq-snr-perclass} reports the Spearman correlation between predicted uncertainty and SNR on the manually curated evaluation set, stratified by predictive outcome. The strongest negative correlation is observed for true positives, indicating that uncertainty for correctly identified transient detections is strongly linked to signal quality. For true negatives, the correlation is weaker, suggesting that uncertainty on bogus detections depends less directly on photometric SNR and more on heterogeneous artifact morphology. Correlations for false positives and false negatives are small and unstable in sign, indicating that misclassifications are not explained by SNR alone.

\begin{table}[h]
\centering
\small
\setlength{\tabcolsep}{6pt}
\renewcommand{\arraystretch}{1.2}
\begin{tabular}{l|cccc}
\hline
$\boldsymbol{\rho}$ \textbf{SNR} & TP & FP & TN & FN \\
\hline
MC Dropout           & $-0.368$ & $0.089$ & $-0.157 $ & $0.063$ \\
Deep Ensemble        & $-0.434 $ & $-0.028 $ & $-0.192 $ & $-0.105$  \\
Repulsive Ensemble   & $-0.420 $ & $0.084$ & $-0.162$ & $0.123$ \\
\textbf{Our method}  & $-0.549$ & $-0.038$ & $-0.166$ & $0.108$ \\
\hline
\end{tabular}
\caption{Spearman rank correlation coefficients ($\rho$) between predicted uncertainty and signal-to-noise ratio (SNR) for each predictive class, True Positive (TP), False Positive (FP), True Negative (TN) and False Negative (FN).}
\label{tab:uq-snr-perclass}
\end{table}

\section{Uncertainty Quantification formalism}
\label{app:uq-formalism}

\subsection{Deep Ensemble}
During inference, the ensemble prediction is obtained by averaging the individual model probabilities:
\begin{equation}
\bar{p}(\mathbf{x}) = \frac{1}{M} \sum_{m=1}^{M} \sigma(f(\mathbf{x; \hat\theta_m})),
\end{equation}
where $\sigma(\cdot)$ denotes the sigmoid activation function, and $\hat\theta_m$ the trained weights of the $m$-th model. The final predicted class label is then given by:
\begin{equation}
\hat{y}_{\text{ensemble}} = \mathbb{I}[\bar{p}(\mathbf{x}) > 0.5],
\end{equation}
with $\mathbb{I}[\cdot]$ representing the indicator function.

Beyond improving prediction stability, the ensemble also facilitates uncertainty quantification. The predictive uncertainty for a given input $\mathbf{x}$ is estimated as the variance of the individual model probabilities:
\begin{equation}
\mathrm{Var}_{\text{ens}}(\mathbf{x}) = 
\frac{1}{M} \sum_{m=1}^{M} 
\big(\sigma(f_m(\mathbf{x})) - \bar{p}(\mathbf{x})\big)^2.
\end{equation}
This measure reflects the degree of agreement among ensemble members, with higher variance indicating greater predictive uncertainty.
\subsection{Repulsive ensemble}

Operationally, the interaction term of the repulsive ensemble is encoded through an update of the form
\begin{equation}
    \phi(f_i^t) = \nabla_{f_i^t}\log p(f_i^t|D)\;-\;\mathcal{R}\Big(\big\{\nabla_{f_i^t}k(f_i^t,f_j^t)\big\}_{j=1}^{M}\Big),
\end{equation}
where the first term promotes agreement with the data while the second term introduces repulsion via a kernel $k(\cdot,\cdot)$ measuring similarity between ensemble functions. The operator $\mathcal{R}(\cdot)$ aggregates the repulsive contributions across the other ensemble members, pushing $f_i$ away from functions that are already represented in the ensemble.

This repulsivity is motivated by the desire of the uncertainty quantification method to cover a larger area of the loss landscape and avoid ensembles that are accurate but overly concentrated around a single mode. At inference time, the procedure remains the same as for Deep Ensembles: predictions are obtained by aggregating the ensemble outputs, and uncertainty can be quantified from the dispersion across members.

\subsection{MC Dropout}
Let  $\hat\theta$ be the trained weights. Given $N$ dropout masks, we sample the dropout masks $z_n \sim q(z)$ (Bernoulli masks based on the dropout rate of the training), so that for each input $\mathbf{x}^*$ we perform $N$ forward passes with these different masks.
For pass $n$, let $f_n(\mathbf{x}^*; \hat\theta,z_n )$ denote the pre-sigmoid logit
of the network. The corresponding predictive probability is

\begin{equation}
    p_n(\mathbf{x}^*) = p(y^* = 1 \mid \mathbf{x}^*, \hat\theta, z_n)
    = \sigma\big(f_n(\mathbf{x}^*; \hat\theta, z_n)\big),
\end{equation}

with $\sigma$ the sigmoid function of the last layer.

The MC dropout predictive probability is then approximated by the first moment:
\begin{equation}
    \bar{p}(\mathbf{x}^*) = \frac{1}{N} \sum_{n=1}^{N} p_n(\mathbf{x}^*)
    = \frac{1}{N} \sum_{n=1}^{N} \sigma\big(f_n(\mathbf{x}^*; \hat\theta, z_n)\big).
\end{equation}

The (epistemic) predictive variance can be approximated by the second centered moment:
\begin{equation}
\mathrm{Var}_{\text{MCdropout}}(\mathbf{x}^*) =
\frac{1}{N} \sum_{n=1}^{N}
\big(p_n(\mathbf{x}^*) - \bar{p}(\mathbf{x}^*)\big)^2.
\end{equation}

\section{UMAP representation}
\label{app:UMAP-overlay}

\begin{figure}[h]
    \centering
    \includegraphics[width=1\linewidth]{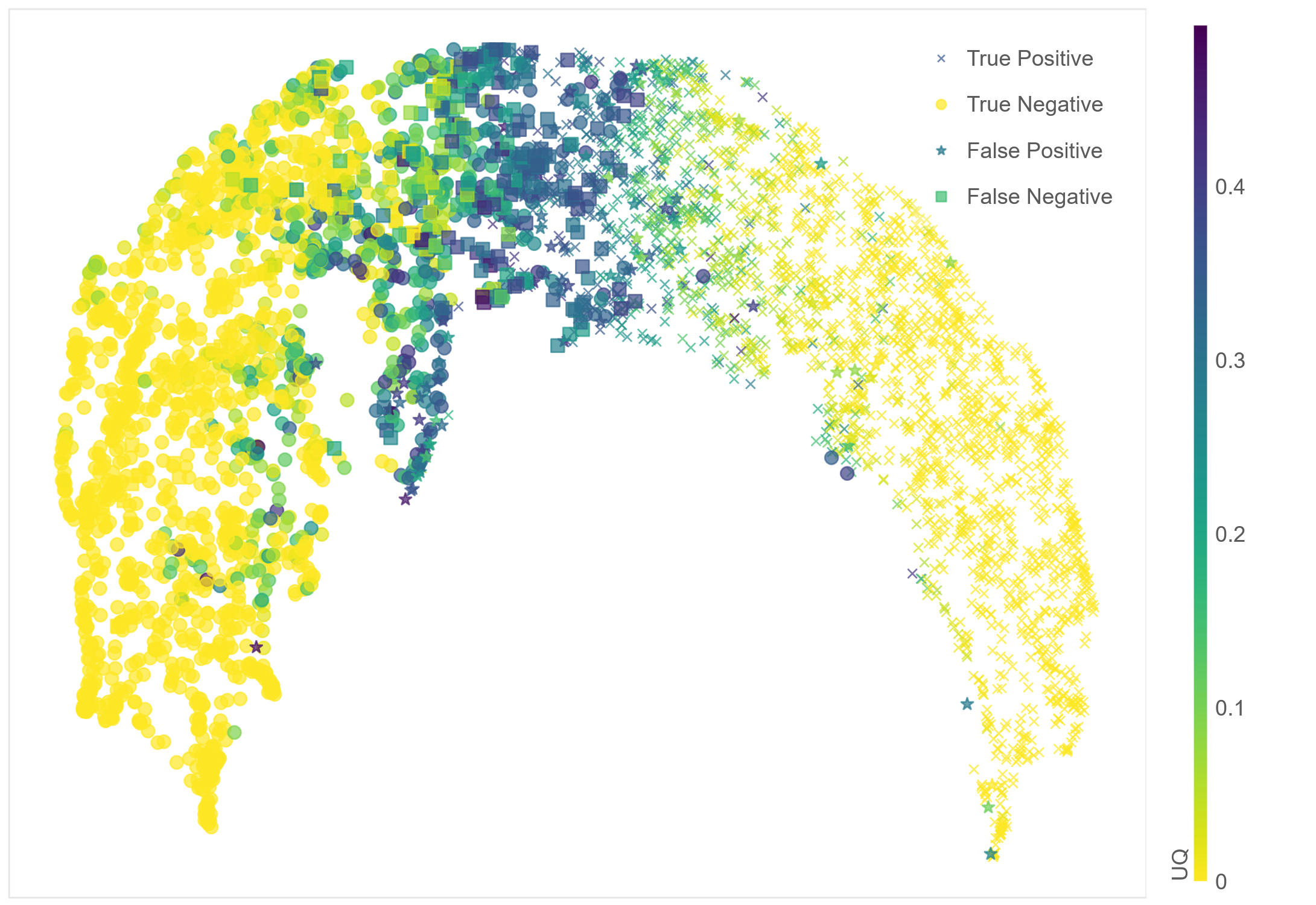}
    \caption{Static UMAP representation of the latent space with predictive uncertainty shown as a color overlay. Symbols represent model predictions. The manually curated evaluation set is shown, and uncertainty is computed using the method introduced in Section~\ref{sec:unc_asy_coteaching}. The displayed uncertainty is the per-sample standard deviation of the predicted probability.}    
    \label{fig:UMAP-UQ}
\end{figure}

\begin{figure}[h]
    \centering
    \includegraphics[width=1\linewidth]{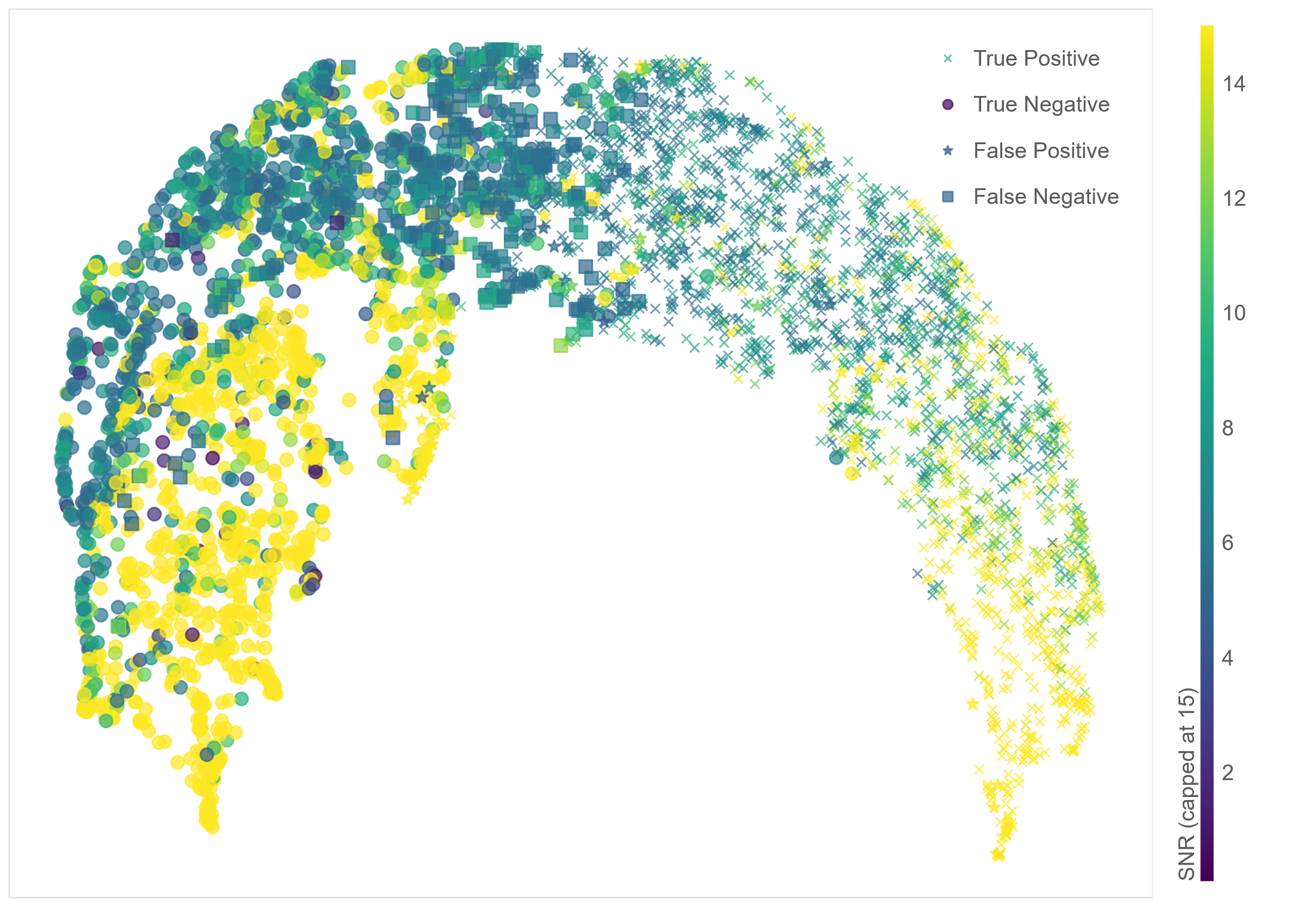}
    \caption{Static UMAP representation of the latent space with signal-to-noise ratio shown as a color overlay. Values are capped at $\mathrm{SNR}=15$. Symbols represent model predictions. The manually curated evaluation set is shown; accordingly, no displayed sources have $\mathrm{SNR}<5$.}
  \label{fig:UMAP_SNR}
\end{figure}

The UMAP projection can be overlaid with additional quantities such as predictive uncertainty and signal-to-noise ratio. These overlays provide a qualitative view of how uncertainty and SNR are distributed across the learned representation.
Figure~\ref{fig:UMAP-UQ} shows that high-uncertainty samples concentrate near the interface between the main regions of the embedding. Figure~\ref{fig:UMAP_SNR} shows that SNR only partially follows the same structure, with a clearer trend in the transient-like region than in the bogus region. These visual patterns are consistent with the quantitative results reported in Section~\ref{sec:results-uq}.

\section{Cumulative light-curve counts}
\label{app:cumul-lightcurve-count}
\begin{table*}[ht]

\centering
\begin{tabular}{r r r r r r}
\toprule
\textbf{Threshold (\%)} & \textbf{All LCs} & \textbf{SN} & \textbf{other transients or variables} & \textbf{Transients (all)} & \textbf{Bogus} \\
\midrule
$\le$   5 & 174 & 33 (100.00\%) & 41 (100.00\%) & 74 (100.00\%) & 174 ( 75.00\%) \\
$\le$  10 & 190 & 33 (100.00\%) & 41 (100.00\%) & 74 (100.00\%) & 190 ( 81.90\%) \\
$\le$  15 & 207 & 33 (100.00\%) & 41 (100.00\%) & 74 (100.00\%) & 207 ( 89.22\%) \\
$\le$  20 & 215 & 33 (100.00\%) & 41 (100.00\%) & 74 (100.00\%) & 215 ( 92.67\%) \\
$\le$  25 & 222 & 33 (100.00\%) & 41 (100.00\%) & 74 (100.00\%) & 222 ( 95.69\%) \\
$\le$  30 & 223 & 33 (100.00\%) & 41 (100.00\%) & 74 (100.00\%) & 223 ( 96.12\%) \\
$\le$  35 & 227 & 33 (100.00\%) & 41 (100.00\%) & 74 (100.00\%) & 227 ( 97.84\%) \\
$\le$  40 & 231 & 33 (100.00\%) & 40 ( 97.56\%) & 73 ( 98.65\%) & 230 ( 99.14\%) \\
$\le$  45 & 232 & 33 (100.00\%) & 40 ( 97.56\%) & 73 ( 98.65\%) & 231 ( 99.57\%) \\
$\le$  50 & 232 & 33 (100.00\%) & 40 ( 97.56\%) & 73 ( 98.65\%) & 231 ( 99.57\%) \\
$\le$  55 & 232 & 33 (100.00\%) & 40 ( 97.56\%) & 73 ( 98.65\%) & 231 ( 99.57\%) \\
$\le$  60 & 233 & 33 (100.00\%) & 40 ( 97.56\%) & 73 ( 98.65\%) & 232 (100.00\%) \\
$\le$  65 & 238 & 30 ( 90.91\%) & 39 ( 95.12\%) & 69 ( 93.24\%) & 232 (100.00\%) \\
$\le$  70 & 243 & 28 ( 84.85\%) & 35 ( 85.37\%) & 63 ( 85.14\%) & 232 (100.00\%) \\
$\le$  75 & 244 & 28 ( 84.85\%) & 34 ( 82.93\%) & 62 ( 83.78\%) & 232 (100.00\%) \\
$\le$  80 & 248 & 26 ( 78.79\%) & 33 ( 80.49\%) & 59 ( 79.73\%) & 232 (100.00\%) \\
$\le$  85 & 256 & 22 ( 66.67\%) & 28 ( 68.29\%) & 50 ( 67.57\%) & 232 (100.00\%) \\
$\le$  90 & 265 & 19 ( 57.58\%) & 23 ( 56.10\%) & 42 ( 56.76\%) & 232 (100.00\%) \\
$\le$  95 & 279 & 11 ( 33.33\%) & 16 ( 39.02\%) & 27 ( 36.49\%) & 232 (100.00\%) \\
$\le$ 100 & 306 & 3 (  9.09\%) & 8 ( 19.51\%) & 11 ( 14.86\%) & 232 (100.00\%) \\
\bottomrule
\end{tabular}
\caption{Cumulative number of light-curves below a predicted-transient threshold (All LCs, Bogus) and above the bin lower edge for transient subclasses (SN-like, other transients or variables, Transients all). Percentages are cumulative within each class/subclass.}
\label{tab:lc_id_phase_summary}
\end{table*}

\end{appendix}

\end{document}